\documentclass[cup6b]{cupbook_mod}

\usepackage{epsf}
\usepackage{graphicx}
\usepackage{multirow}

\title{The Formation and Evolution\\
 of Prestellar Cores}
\def\kms{\ifmmode {\,{\rm km\,s^{-1}}}	% km s-1
        \else {\hbox{$\,$ {\rm km$\,$s$^{\rm -1}$}}}\fi}

\def\ltsima{$\; \buildrel < \over \sim \;$}
\def\simlt{\lower.5ex\hbox{\ltsima}}
\def\gtsima{$\; \buildrel > \over \sim \;$}
\def\simgt{\lower.5ex\hbox{\gtsima}}
\def\h2{\mbox{$_{\mbox{\tiny H2}}$}}
\def\k13{\mbox{$\kappa_{\mbox{\tiny 1.3}}$}}
\def\I13{\mbox{$I_{\mbox{\tiny 1.3}}$}}

\def\cm2{\mbox{$\mbox{cm}^{-2}$}}
\def\cm3{\mbox{$\mbox{cm}^{-3}$}}

\def\h2{\mbox{$_{\mbox{\tiny H2}}$}}
\def\k13{\mbox{$\kappa_{\mbox{\tiny 1.3}}$}}
\def\I13{\mbox{$I_{\mbox{\tiny 1.3}}$}}

\def\msun{\mbox{M$_\odot$}}
\newcommand{\kmps}{\,{\rm km\, s^{-1}} }

\newcommand{\vect}[1]{\mbox{\boldmath$#1$}}

\newcommand{\ceta}{c_{\eta}}

\newcommand{\cmci}{\,{\rm cm}^{-3} }

\newcommand{\dfrac}[2]{{\displaystyle \frac{#1}{#2}}  }

\newcommand{\cs}{c_{\rm s}}
\newcommand{\csq}{c_{\rm s}^2}

\begin{document}

%\author[Ph. Andr\'e, S. Basu, and S. Inutsuka]{PHILIPPE ANDRE$^1$,
%SHANTANU BASU$^2$, and SHU-ICHIRO INUTSUKA$^{3}$\\
%(1) Service d'Astrophysique, UMR AIM, CEA/DSM - CNRS - Univ. Paris 7, CEA Saclay, France\\
%(2) Department of Physics and Astronomy, Univ. of Western Ontario, London, Canada\\
%(3) Department of Physics, Graduate School of Science, Kyoto Univ., Japan\\
%}

\bigskip
\bigskip

\author{
Philippe Andr\'e, Service d'Astrophysique, UMR AIM,\\ 
CEA/DSM - CNRS - Univ. Paris 7, CEA Saclay, France\\
\medskip
Shantanu Basu, Department of Physics and Astronomy,\\ 
University of Western Ontario, London, Canada\\
\medskip
Shu-ichiro Inutsuka, Department of Physics,\\ 
Graduate School of Science, Kyoto University, Japan}

\date{ }

\maketitle

%\chapter{The Formation and Evolution of Prestellar Cores}

\bigskip
\bigskip
\bigskip
\bigskip

\noindent
\centerline{{\Large {\bf Abstract}}}

\medskip
\noindent

Improving our understanding of the initial conditions and earliest stages 
of star formation is crucial to gain insight into the origin of stellar 
masses, multiple systems, and protoplanetary disks. We review the properties 
of low-mass dense cores as derived from recent millimeter/submillimeter 
observations of nearby molecular clouds and discuss them in the context of 
various contemporary scenarios for cloud core formation and evolution.
None of the extreme scenarios can explain all observations. Pure laminar 
ambipolar diffusion has relatively long growth times for typical ionization 
levels and has difficulty satisfying core lifetime constraints. Purely 
hydrodynamic pictures have trouble accounting for the inefficiency of core 
formation and the detailed velocity structure of individual cores. A possible 
favorable scenario is a mixed model involving gravitational fragmentation of 
turbulent molecular clouds close to magnetic criticality. The evolution of the 
magnetic field and angular momentum in individual cloud cores after the onset 
of gravitational collapse is also discussed. In particular, we stress the 
importance of radiation-magnetohydrodynamical processes and resistive MHD 
effects during the protostellar phase. We also emphasize the role of the 
formation of the short-lived first (protostellar) core in providing a chance 
for sub-fragmentation into binary systems and triggering MHD outflows. Future 
submillimeter facilities such as Herschel and ALMA will soon provide major new 
observational constraints in this field. On the theoretical side, an important 
challenge for the future will be to link the formation of molecular clouds and 
prestellar cores in a coherent picture.  

\newpage

\section{Introduction: Dense cores and the origin of the IMF}
\label{intro}

Stars form from the gravitational collapse of dense cloud cores in the molecular interstellar medium 
of galaxies.  Studying and characterizing the properties of  dense cores is thus of great interest 
to gain insight into the initial conditions and initial stages of the star formation process.

Our observational understanding of low-mass dense cores has made significant progress in 
recent years and three broad categories of cores can now be distinguished within nearby 
molecular clouds, which possibly represent an evolutionary sequence: starless cores, prestellar 
cores, and ``Class~0'' protostellar cores.
Starless cores are possibly transient concentrations of molecular gas and dust 
without embedded young stellar objects (YSOs), typically observed in tracers such as
C$^{18}$O (e.g. Onishi et al. 1998), NH$_{3}$ (e.g. Jijina, Myers, \& Adams 1999), or 
dust extinction (e.g. Alves et al. 2007), and which do not show evidence of infall.
Prestellar cores are also starless ($M_\star = 0$) but represent a somewhat denser and more 
centrally-concentrated population of cores which are self-gravitating, hence unlikely 
to be transient. They are typically detected in (sub)millimeter dust continuum emission and  
dense molecular gas tracers such as NH$_3$ or N$_2$H$^+$ 
(e.g. Ward-Thompson et al. 1994; Benson \& Myers 1989; Caselli et al. 2002), 
often seen in absorption at mid- to far-infrared wavelengths 
(e.g. Bacmann et al. 2000, Alves et al. 2001), 
and frequently exhibit evidence of infall motions (e.g. Gregersen \& Evans 2000).
Conceptually, all prestellar cores are starless but only a subset of the starless cores evolve into
prestellar cores; the rest are presumably ``failed'' cores that eventually disperse and never 
form stars.  In practice, prestellar cores are characterized by large density contrasts over the local 
background medium. Specifically, the mean densities of observed prestellar cores  
exceed the mean densities of their parent clouds by a factor $\simgt $~5--10,  
while their mean column densities exceed the background column densities by a factor $\simgt $~2.
For comparison, a critical self-gravitating Bonnor-Ebert isothermal spheroid has a mean density 
contrast $\bar{\rho}_{\rm BE}/\rho_{\rm ext} \sim 2.4$ (e.g. Lombardi \& Bertin 2001)  
and a mean column density contrast $\bar{\Sigma}_{\rm BE}/\Sigma_{\rm ext} \sim 1.5$ 
over the external medium.

Finally, Class~0 cores/objects are young accreting protostars observed early after point mass
formation while most of the mass of the system is still in the form of a dense core/envelope 
as opposed to a YSO ($M_\star \ll M_{\rm env} $) (Andr\'e, Ward-Thompson, Barsony 1993).
They are believed to result from the gravitational collapse of prestellar cores.
Class~0 protostars themselves evolve into Class~I objects with $M_{\rm env} < M_\star$ 
(Lada 1987;  Andr\'e \& Montmerle 1994) as the protostellar envelope dissipates 
through accretion and ejection of circumstellar material.  Class~I objects subsequently 
evolve into (Class~II and Class~III)  pre-main sequence stars surrounded by a circumstellar disk 
(optically thick and optically thin in the near-/mid-IR, respectively), 
but lacking a dense circumstellar envelope ($M_{\rm env} \sim 0$). 

Improving our understanding of the formation and evolution of dense cores in 
molecular clouds  
is crucially important since there is now good evidence that these early stages 
largely control the origin of the stellar initial mass function (IMF).
Indeed, observations indicate that the prestellar core mass function resembles 
the IMF (see \S ~\ref{sec:surveys} below),  suggesting that the effective reservoirs 
of mass required for the formation of individual stars are already
selected at the prestellar core stage. 
Furthermore, it is at the very end of the prestellar stage that multiple systems are 
believed to form and during the protostellar stage that a fraction of the prestellar core mass 
reservoir is accreted by the central protostellar components as a result of the accretion/ejection
process.

\begin{figure}
\centering
 \includegraphics[width=8cm,angle=270]{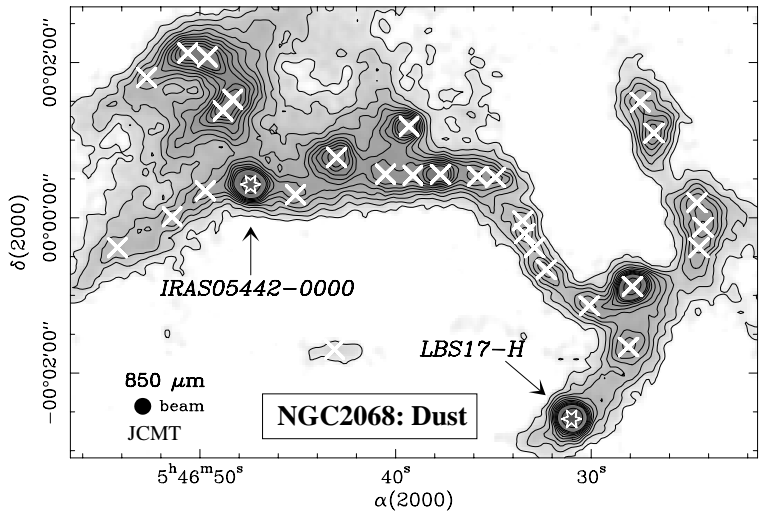}
  \caption{SCUBA $850\:\mu$m dust continuum map of the NGC~2068 protocluster
extracted from the mosaic 
of Orion~B by Motte et al. (2001).  
A total of 30 compact prestellar condensations (marked by crosses), with masses 
between $\sim 0.4\, M_\odot $ and $\sim 4.5\, M_\odot $, are detected 
in this $\sim 1$~pc $\times$ 0.7~pc field.}
\label{fig:ngc2068}
\end{figure}

\section{Link between the prestellar core mass function and the IMF}
\label{sec:surveys}

Wide-field (sub)mm dust continuum mapping is a powerful tool to 
take a census of prestellar dense cores and young protostars 
within star-forming clouds. 
The advent of large-format bolometer arrays on (sub)millimeter 
radiotelescopes such as the IRAM 30m and the JCMT has led to the identification
of numerous cold, compact condensations that do not obey the 
Larson (1981) self-similar scaling laws of molecular clouds 
and are intermediate in their properties 
between diffuse CO clumps and infrared young stellar objects  
(cf. Andr\'e et al. 2000 and Ward-Thompson et al. 2007 for reviews). 
As an example, Fig.~\ref{fig:ngc2068} shows 
the condensations found by Motte et al. (2001) at 850~$\mu$m in the 
NGC~2068 protocluster (Orion~B). 
Such highly concentrated (sub)millimeter continuum condensations are 
at least 3 to 6 orders of magnitude denser than typical CO clumps 
(e.g. Kramer et al. 1998) and feature large ($\gg 50 \%$) mean column 
density contrasts over their parent 
background clouds, strongly suggesting they are self-gravitating. 
The latter is directly confirmed by line observations in a number of cases.
When available, the virial masses of the condensations indeed agree within a factor of $\sim 2$
with the masses derived from the (sub)millimeter dust continuum 
(e.g. Andr\'e et al. 2007). 
A small fraction of these condensations lie at the base of powerful jet-like 
outflows and correspond to Class~0 objects.
However, the majority of them are starless/jetless  
and appear to be the immediate prestellar 
progenitors of individual protostars or protostellar systems. 

\begin{figure}
\centering
\includegraphics[width=8cm,angle=270]{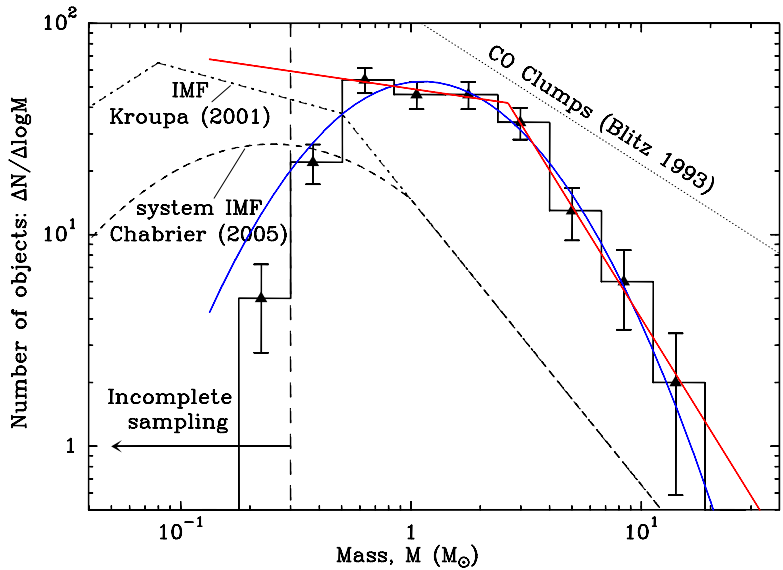}
\caption{Differential (dN/dlogM) mass distribution of the 229 starless dust continuum 
condensations detected at 850~$\mu$m with SCUBA in the Orion~A/B cloud complex 
excluding the crowded OMC1 and NGC~2024 regions 
(histogram with error bars -- from Motte et al. 2001, 
Johnstone et al. 2001, and Nutter \& Ward-Thompson 2007). 
This prestellar core sample is estimated to be complete down to $\sim 0.3\, M_\odot $.
A two-segment power law fit and a lognormal fit are shown for comparison. 
The lognormal fit peaks at $\sim 1.1\, M_\odot $ and has a standard deviation of $\sim 0.41$ 
in log$_{10}$M.
For reference, the dash-dotted curve shows the shape of the single-star IMF  (e.g. Kroupa 2001)  
and  the dashed curve corresponds to the IMF of multiple systems (e.g. Chabrier 2005).  
(The lognormal part of the latter peaks at $0.25\, M_\odot $ and has a standard deviation of 
0.55 in log$_{10}$M.)
The dotted line shows a $\rm{dN/dlogM} \propto \rm{M}^{-0.6}$ power-law 
distribution corresponding to the typical mass spectrum found 
for low-density CO clumps (see Blitz 1993 and Kramer et al. 1998).  
}
\label{fig:orion_cmf}
\end{figure}

In particular, as first pointed out by Motte, Andr\'e, Neri (1998) in 
the case of the $\rho $~Ophiuchi (L1688) cloud,  
the mass distribution of these starless dust continuum condensations  
is remarkably similar in shape to the stellar IMF (see Fig.~\ref{fig:orion_cmf}).
This was consistently found by a 
number of independent groups in the past few years (e.g. Testi \& Sargent 1998;
Johnstone et al. 2000, 2001; Motte et al. 2001; Stanke et al. 2006; Enoch et al. 2006; 
Nutter \& Ward-Thompson 2007) in nearby star-forming regions such as $\rho $~Ophiuchi, Serpens, Orion A \& B, 
and Perseus. 
In all of these clouds, the observed prestellar core mass function (CMF) is consistent with the 
Salpeter (1955) power-law IMF at the high-mass end ($\rm{dN/dlogM} \propto \rm{M}^{-1.35}$), 
and thus significantly steeper than the mass distribution of diffuse CO clumps 
($\rm{dN/dlogM} \propto \rm{M}^{-0.6}$ over at least three decades in mass -- e.g. Blitz 1993; Kramer et al. 1998).
The difference presumably arises because CO clumps are primarily structured
by supersonic turbulence (e.g. Elmegreen \& Falgarone 1996) while prestellar condensations 
are largely free of supersonic turbulence and clearly shaped by self-gravity 
(e.g. Motte et al. 2001, Andr\'e et al. 2007 and \S~\ref{observations} below). 
The slope of the observed CMF becomes shallower than the Salpeter power law 
and more similar to the slope of the typical CO clump mass distribution at the low-mass end.
Based on the results of present core surveys, the entire prestellar CMF can generally be fit equally 
well with either a two-segment broken power law or a lognormal distribution down to the completeness 
limit of the observations.
This is illustrated in Fig.~\ref{fig:orion_cmf} which shows the differential CMF in log-log format for a sample of 
229 starless submillimeter continuum cores in the Orion A/B complex identified in the SCUBA surveys of 
Motte et al. (2001) and Johnstone et al. (2001, 2006). 

Note that there is some discussion in the literature (e.g. Reid \& Wilson 2006) as to whether the differential 
or the cumulative form of the CMF should be used. The differential form is more intuitive to interpret 
as it can be fit using the standard least-squares technique after assigning simple Poisson error bars.
It is however inadequate when the total number of objects is small ($\simlt 100$) as it is significantly
affected by the arbitrary choice of mass bins. The cumulative form, which is independent of binning,  
is preferable when dealing with small samples, but some care must be taken when comparing it 
with model distributions based on least-squares fits (cf. Reid \& Wilson 2006). 
In practice, using the non-parametric 
Kolmogorov-Smirnov (K-S) test to compare the cumulative form of the CMF with model distributions 
is  more reliable than least-squares fitting.
The K-S test confirms that, within statistical uncertainties, the prestellar CMF observed in the 
above-mentioned nearby clouds is indistinguishable in shape from the stellar IMF, although the significance 
remains limited by the relatively small size of present prestellar core samples. 
The K-S test also shows that the observed CMF differs from the shallow power-law mass distribution 
of CO clumps at a very high significance level (for instance, in the Orion case illustrated in Fig.~\ref{fig:orion_cmf}, 
the probability that the two mass distributions are statistically similar in shape is 
only $P \approx 2\times 10^{-6} $).

In addition, the median prestellar core mass observed in regions such as 
$\rho$~Ophiuchi and Orion ($\sim 0.2-1.5\, M_\odot $) is only slightly larger than the characteristic 
$\sim 0.5\, M_\odot $ set by the peak of the IMF in dN/dlogM format. 
Such a close resemblance of the CMF to the IMF 
in both shape and mass scale is consistent with the view that the prestellar  
condensations identified in (sub)millimeter dust continuum surveys 
are about to form stars on a one-to-one basis, with a fixed and relatively 
high local efficiency, i.e., $\epsilon_{\rm core} \equiv M_\star /M_{\rm core} \simgt 30-50\%$. 

Interestingly, in a recent near-IR extinction imaging study of the Pipe dark cloud, 
Alves et al. (2007) found a population of 159 starless cores whose mass distribution 
similarly follows the shape of the IMF. This finding is reminiscent of the CMF results 
obtained from the (sub)millimeter dust continuum, although it is important to stress that most of the 
starless cores in the Pipe Nebula are gravitationally unbound objects confined by external 
pressure (Lada et al. 2008). Hence, they do not qualify as prestellar cores and a large fraction 
of them may never evolve into stars. Assuming nevertheless that most of them will evolve 
into self-gravitating prestellar cores and subsequently collapse into stars, the 
Alves et al. (2007) result suggests that the IMF may be determined even earlier than 
the prestellar stage.

Appealing as a direct connection between the prestellar CMF 
and the IMF might be, several caveats should be kept in mind. 
First, although core mass estimates based on optically thin 
(sub)millimeter dust continuum emission are straightforward, 
they rely on uncertain assumptions about the 
{\it dust (temperature and emissivity) properties} (e.g. Stamatellos et al. 2007a).
Second, current determinations of the CMF are limited by 
small-number statistics in any given cloud and may be affected 
by incompleteness at the low-mass end (e.g. Johnstone et al. 2000).
With $Herschel$, the future submillimeter space telescope to be launched 
in 2008, it will be possible to dramatically improve on the statistics 
and to largely eliminate the mass uncertainties through direct 
measurements of  the dust temperatures (cf.  Andr\'e \& Saraceno 2005). 
Third, in some regions such as the $\rho $ Ophiuchi cloud, 
the shape of the CMF is in better agreement with the 
IMF of individual field stars than with the IMF of multiple systems 
(e.g. Andr\'e et al. 2007). 
This is surprising since the spatial resolution of current surveys for prestellar 
cores ($\sim 2000$~AU at best) is not sufficient to probe core multiplicity. 
Furthermore, multiple systems are believed to form {\it after} the prestellar 
stage by subsequent dynamical fragmentation during the collapse phase, 
close to the time of protostar formation (e.g. Goodwin et al. 2007).
Thus, one would expect the masses of prestellar cores to be more 
directly related to the masses of multiple systems than to the masses of 
individual stars. It is possible that a fraction at least of the cores observed 
at masses lower than the peak of the CMF in dN/dlogM format  
(e.g.,  $\sim 1.1\, M_\odot $ in Fig.~\ref{fig:orion_cmf})  
are not gravitationally bound, hence not prestellar in nature 
(cf. Andr\'e et al. 2007). 

Last but not least, there is a potential timescale problem.
As pointed out by Clark, Klessen, \& Bonnell (2007), 
if the lifetime of prestellar cores depends 
on their mass, then the observed mass distribution is not necessarily representative of the intrinsic 
CMF (see also Elmegreen 2000). 
This is due to the fact that an observer is more likely to detect long-lived cores than short-lived cores. 
In practice, however, the mean densities of prestellar cores are essentially uncorrelated 
with their masses, so that there is no systematic dependence of the dynamical timescale on the mass.
The importance of the potential timescale bias can be assessed by considering a {\it weighted} 
core mass function in which each core is assigned a weight equal to 
$\frac{\langle t_{\rm ff} \rangle}{t_{\rm ff}} = \frac{\bar{\rho}^{1/2}}{\langle\bar{\rho}^{1/2}\rangle}$ (instead of 1), 
where $\langle t_{\rm ff}\rangle$ is the average free-fall time of the sampled cores. 
Such a weighting makes it possible to recover the intrinsic shape of the CMF if the 
lifetime of each core is proportional to its free-fall time.  
Andr\'e et al. (2007) applied this technique to 
the sample of 57 starless dust continuum 
condensations identified by Motte et al. (1998) at 1.2~mm in the $\rho$~Ophiuchi cloud.
In this case, the  above-mentioned weighting does not change the high-mass end 
of the CMF and only affects the low-mass end: it renders the entire $\rho$~Oph CMF 
derived above the $\sim 0.1\, M_\odot $ completeness level remarkably consistent with 
a single, Salpeter power-law mass distribution (see Fig.~8 of Andr\'e et al. 2007).
We conclude that the steep, Salpeter-like slope of the CMF at the high-mass end is robust, 
but that the departure from a single power-law distribution at the low-mass end, in the form of  
a break near the median core mass (cf. Fig.~\ref{fig:orion_cmf}),  
is less robust.

Despite these limitations, the observational findings summarized in this 
section are very encouraging as they support scenarios according to which  
the bulk of {\it the IMF is partly determined by pre-collapse cloud 
fragmentation} (e.g. Larson 1985, 2005; Elmegreen 1997; 
Padoan \& Nordlund 2002). The finding that the high-mass end of the prestellar CMF
is substantially steeper than the $\rm{dN/dlogM} \propto \rm{M}^{-0.6}$ mass distribution 
of low-density CO clumps is very significant: while most of the mass of the low-density 
CO medium is contained in the largest, most massive clumps, most of the prestellar mass 
destined to evolve into stars is in small, low-mass cores. 
Clearly, one of the keys to the problem of the origin of the IMF 
lies in a good understanding of the processes responsible for the formation 
of prestellar cores/condensations out of low-density structures within molecular clouds.
However, it is likely that additional processes, such as subfragmentation 
into binary/multiple systems and more generally mechanisms controlling the 
value of the star formation efficiency at the core level ($\epsilon _{\rm core} $), 
also play an important role and, in particular, are required 
to generate the low-mass ($M < 0.3\, M_\odot $) end of the IMF 
(cf. Bate et al. 2003, Ballesteros-Paredes et al. 2006).
In the following sections, we discuss core formation and core subfragmentation 
models in turn.

\section{Core formation models vs. observational constraints}
\label{formation}

The mechanisms by which prestellar cores form and evolve in molecular clouds 
are the subject of a major theoretical debate at the moment. There is little doubt that 
self-gravity ultimately plays a dominant role and it has even been proposed that dense cores 
may form by purely gravitational fragmentation (e.g. Larson 1985;  Hartmann 2002).
However, the respective roles of magnetic fields and interstellar turbulence in regulating 
the core/star formation process are highly controversial. In particular, the 
classical picture of slow, quasi-static core formation by 
ambipolar diffusion in magnetically-supported clouds 
(e.g. Mouschovias 1987; Shu et al. 1987, 2004) 
has been seriously challenged by a new, much more dynamic paradigm, which 
emphasizes the role of supersonic turbulence in supporting clouds on large
scales and generating density fluctuations on small scales 
(e.g. Padoan \& Nordlund 2002; Mac Low \& Klessen 2004).
Conceptually, core formation models may be conveniently divided up into four 
categories, depending on whether linear 
or non-linear (turbulent) perturbations 
initiate core formation, and on whether magnetic fields are dynamically dominant or not  
(see Table~\ref{basutable1} below). In this classification, the ``standard'' 
picture and the new turbulent 
paradigm correspond to two extreme models, 
according to which cores form by magnetically-regulated gravitational fragmentation 
and super-Alfv\'enic turbulent fragmentation, respectively. In all turbulent models, cores 
are initially formed by cloud material compressed by shocks arising from supersonic turbulence.
There are however several versions of the dynamic picture of core/star formation which 
mainly differ in the way they explain the origin of the IMF. In the Padoan \& Nordlund (2002)
scenario, the IMF is almost entirely set by the properties of interstellar turbulence at the prestellar 
stage, while in the alternative model proposed by Bate \& Bonnell (2005), turbulence 
is largely irrelevant for the IMF which originates from competitive accretion and dynamical interactions at 
the protostellar stage. The Klessen \& Burkert (2000) scenario is intermediate between these two 
extremes in that both turbulence and dynamical interactions play a role in shaping the IMF. 
The various turbulent models also differ depending on whether the turbulence is freely decaying 
(e.g. Tilley \& Pudritz 2007) or continuously driven (e.g.  V\'azquez-Semadeni et al. 2005).

\subsection{Theoretical description of cloud fragmentation models}
\label{theory}

The classical problems of fragmentation of a sheet-like layer or
a cylinder are relevant to star formation 
because they show that there is a {\it preferred} scale of gravitational 
fragmentation as soon as one considers a structured (i.e., flattened
or filamentary) region with some scale length $H$ (see Larson 1985). 
This is not the
case for a uniform medium, in which case the Jeans analysis shows that the
largest possible scale has the fastest growth rate. Sheets and filaments
can be easily generated by the formation process of the molecular cloud
itself or by subsequent internal turbulent or gravitational motions.
For isothermal sheets, $H = \csq/(\pi G\Sigma)$, where $\cs$ 
is the isothermal
sound speed, and $\Sigma$ is the column density of the sheet.
For highly flattened sheets, the preferred fragmentation scale is   
$\lambda_m = 2\pi H$, while for a layer with the extended  
isothermal atmosphere calculated by Spitzer (1942), $\lambda_m
= 4.4 \pi H$ (Simon 1965). 
These two length scales likely bracket the 
possibilities for more realistic isothermal non-magnetic sheet-like 
configurations.
The growth time of the fragmentation instability is 
 essentially the dynamical timescale $t_{\rm d} \simeq H/\cs$ for the 
cases described above, and is more generally identified with 
the free-fall timescale 
 $t_{\rm ff} \simeq 1/\sqrt{G\rho}$ that applies to both unpressured
sheets and those confined by a strong external pressure.

However, purely gravitational fragmentation instability of an entire molecular cloud 
on the dynamical timescale given by the mean column density and temperature
is ruled out (Zuckerman and Palmer 1974), due to the observed low
efficiency of Galactic star formation.
Nevertheless, it remains a relevant concept to understanding fragmentation
in the cluster-forming subregions of molecular clouds.
A proposed explanation for the low efficiency of star formation is the 
so-called ``standard model'' of star formation
(Shu, Adams, \& Lizano 1987; Mouschovias 1987), in which cores are formed
on a diffusive timescale (much longer than than a dynamical timescale), 
due to the ambipolar
drift of neutrals past dynamically dominant magnetic fields. This mode
of star formation is often erroneously labeled as ``isolated'' star
formation. In reality, ambipolar diffusion driven core formation is also 
a fragmentation process as surely as its nonmagnetic counterpart - the
difference is primarily in the timescale of evolution. One may however
make the following distinction based on the mass-to-magnetic flux ratio, 
$M/\Phi $, compared to the critical value, $\left(M/\Phi \right)_{\rm crit} $, necessary 
for support against gravitational collapse. 
Molecular cloud envelopes may be 
magnetically dominated, with subcritical to transcritical mass-to-flux ratios, 
i.e., $\mu_{\rm env} \equiv \left(M/\Phi \right) /  \left(M/\Phi \right)_{\rm crit}  \leq 1$.
Cloud envelopes also have low mean density and relatively high levels of ionization, 
and thereby evolve on such a long timescale for fragmentation that no runaway
has essentially occurred by the time we observe the clouds. 
Conversely, the cluster-forming cores (and also regions of weak clustering found in
the Taurus molecular cloud) may be magnetically supercritical with $\mu_{\rm clus} > 1$ 
and also have greater mean density, resulting in fragment formation on 
relatively shorter timescales and lengthscales.

For the above reasons and because Zeeman measurements (see the compilation
of Crutcher 1999 and \S ~\ref{bfield} below) establish that mass-to-flux ratios are clustered about
the critical value, it is important to study cloud fragmentation including 
the effect of dynamically significant magnetic fields and also ambipolar
diffusion, if possible. Magnetized sheets can be studied most easily
in two limits. If the formation process of the sheet is by dynamical
compression, say from stellar winds or supernovae, the ambient interstellar
magnetic field may be swept up into the expanding shell and be oriented
primarily in the plane of the sheet. Conversely, if self-gravity plays
an important role in the formation of
the sheet, then it will be flattened along the mean direction of the 
magnetic field.

When the magnetic field is in the
plane of the sheet, the character of fragmentation and the fate of
the cloud can be further categorized into two subcases
(Nagai, Inutsuka, \& Miyama 1998).
If the geometrical thickness of the sheet is comparable to (or
larger than) the natural scale height
($\sim c_{\rm s}/\sqrt{G\rho_{\rm c}}$),
or equivalently, if 
the ambient pressure is much smaller than the midplane pressure,
compressional motions along the magnetic field lines result in
the formation of filamentary clouds elongated 
perpendicular to the field lines.
The line-mass (mass per unit length) of the filament is
larger than that of the (isothermal) equilibrium filament
so that collapse toward the axis of the filament continues until 
temperature increases (Inutsuka \& Miyama 1992).
In this case, the characteristic (i.e., minimum) mass scale for 
fragmentation is determined by the final fragmentation of
the filamentary cloud, and significantly smaller than
the initial Jeans mass of the sheet (Inutsuka \& Miyama 1997).
The resulting mass scale can be called ``the minimum Jeans mass'' and
its dependence on the initial temperature and metallicity
of the cloud was discussed by Masunaga \& Inutsuka (1999).
The expected core mass function was analytically derived by 
Inutsuka (2001) using the Press-Schechter formalism.
The overall timescale of the whole process is on the order
of the initial free-fall time ($t_{\rm ff} \sim 1/\sqrt{G\rho_{\rm c}}$).\\
In contrast, if the sheet-like cloud is confined by strong external
pressure due to an ambient warm (or hot) medium, the thickness of
the sheet is smaller than $\sim c_{\rm s}/\sqrt{G\rho_{\rm c}}$.
The timescale for fragmentation is still on the order of the
gravitational free-fall time $t_{\rm ff}$, but sound waves can now propagate
many times across the (thin) sheet within this timescale.
Thus, the fragmentation is in an incompressible mode and
the axes of the resulting filaments are parallel to the direction of 
the magnetic field. In this case, the filaments may fragment into  cores 
whose masses  can be smaller than the Jeans mass.
In effect, this second case provides a mechanism for generating 
gravitationally stable cores (Inutsuka \& Miyama 1997).

\begin{table}
\caption{Summary of main features of core formation models.}
\begin{center}
\begin{tabular}{|l|l|l|}
\hline
\multicolumn{3}{|c|}{Core Formation Scenarios} \\
\hline
Models & Gravitational Fragmentation& Turbulent Fragmentation\\ 
              & (linear perturbations)  & (non-linear perturbations)\\ \hline
\multirow{4}{*}{Weak $B$} & Short (few Myr) timescale & Very short ($<$ Myr) timescale\\
 & infall mildly supersonic & infall highly supersonic\\
 & ordered curved field lines & field lines distorted\\
 & initial CMF very narrow & initial CMF is broad, IMF-like\\ \hline
\multirow{5}{*}{Strong $B$} & Long ($\sim 10$ Myr) timescale & Short (few Myr) timescale\\
 & subsonic infall & subsonic relative infall and\\
 & & supersonic systematic speeds\\ 
 & small field line curvature& ordered field lines\\ 
 & initial CMF very narrow & initial CMF is broad, IMF-like \\ 
\hline
\end{tabular}
\end{center}
\label{basutable1}
\end{table}

When the magnetic field is oriented perpendicular to the sheet,
recent simulations of cloud fragmentation which include ambipolar diffusion 
have allowed an extensive parameter study of the interplay of
magnetic fields, ambipolar diffusion, and turbulence in the formation
of cores (Basu \& Ciolek 2004; Basu, Ciolek, \& Wurster 2008). 
Given the standard cosmic-ray induced ionization fraction
$x_i \simeq 10^{-7}(n/10^4\, {\rm cm}^{-3})^{-1/2}$
(Elmegreen 1979; Nakano 1979),
there is an observationally distinguishable 
difference between subcritical and supercritical fragmentation. 
Subcritical fragmentation leads to subsonic infall motions onto 
cores and within them, while supercritical cloud fragmentation leads
to extended supersonic infall on the core scale $\sim 0.1$ pc and
even beyond.

\begin{figure}
\centering
\begin{tabular}{cc}
\includegraphics[width=5.5cm]{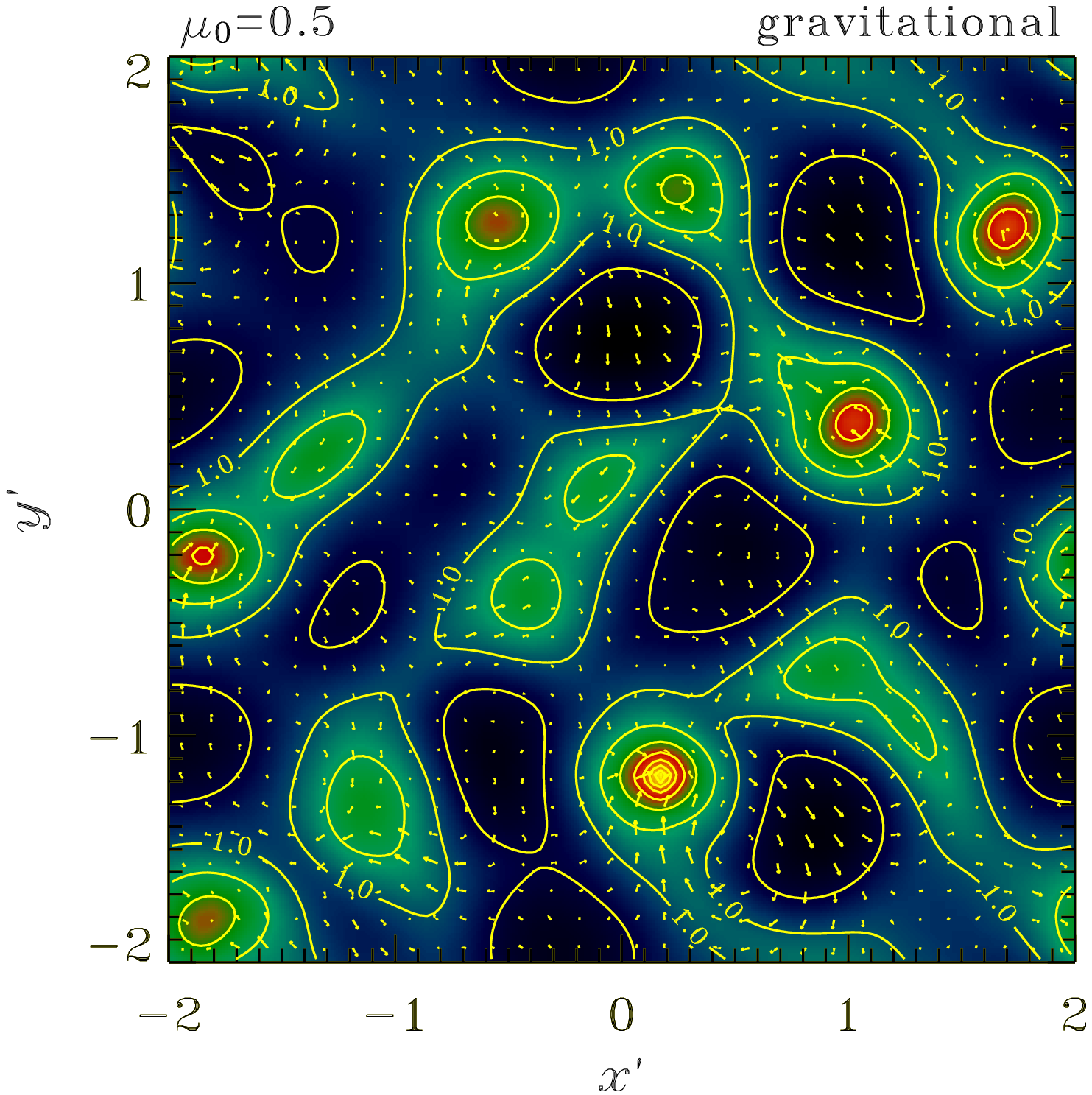} &
\includegraphics[width=5.5cm]{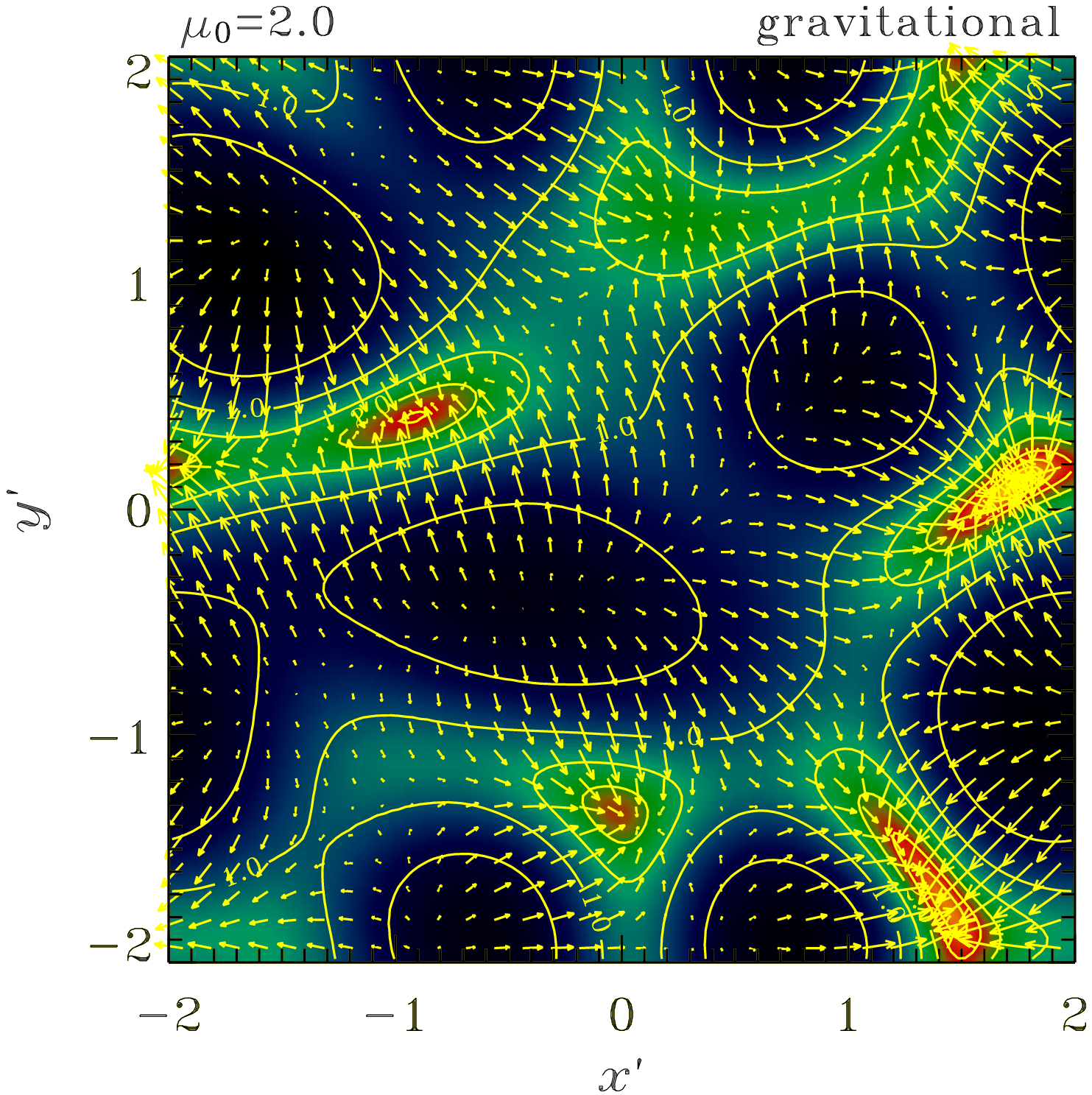}  \\
\includegraphics[width=5.5cm]{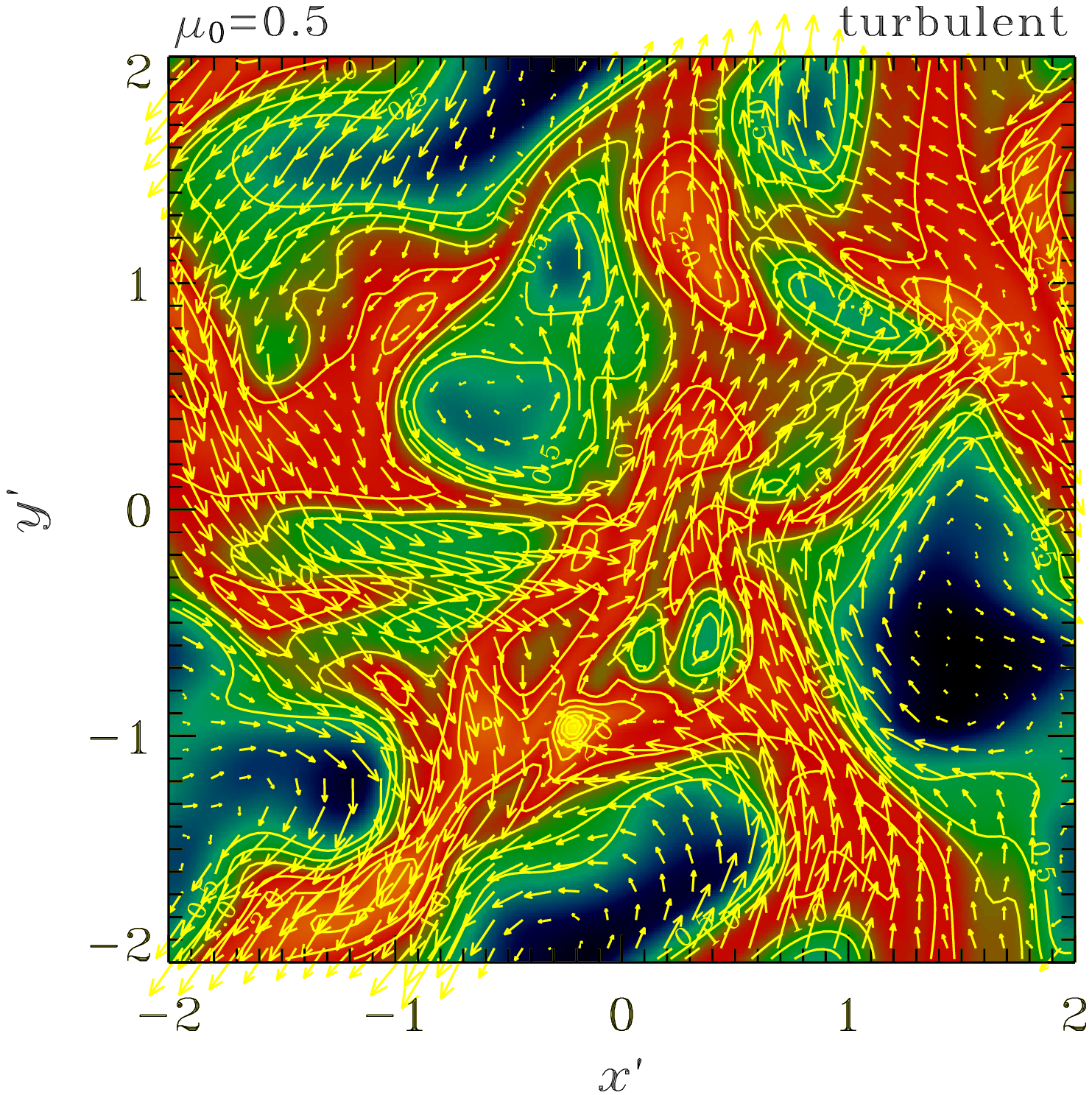} &
\includegraphics[width=5.5cm]{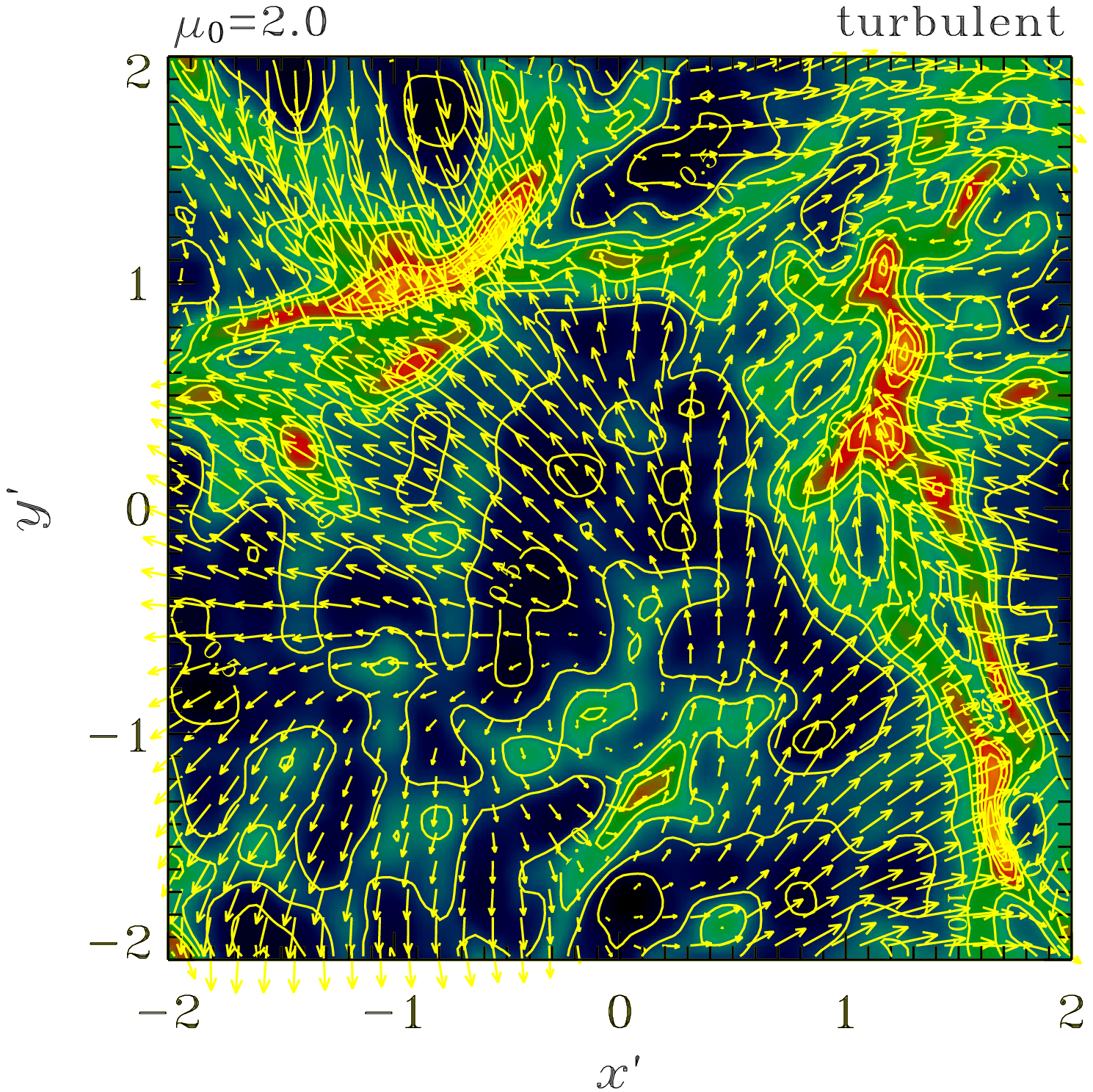} 
\end{tabular}
\caption{Image and contours of column density
and velocity vectors of neutrals, for four different models at the time
of runaway collapse of the first supercritical dense core 
(Basu, Ciolek, \& Wurster 2008; Basu et al. 2008).
Top left: gravitational fragmentation for a subcritical model $\mu_0=0.5$. 
Top right: the same for a supercritical model with $\mu_0=2.0$. 
Bottom left: turbulent fragmentation for $\mu_0=0.5$.
Bottom right: turbulent fragmentation for $\mu_0=2.0$. 
The color table
is applied to the logarithm of column density and the contour lines
represent values of column density enhancement in
multiplicative increments of $2^{1/2}$, having the values
[0.7,1.0,1.4,2,2.8,4.0,...].
The horizontal or vertical distance between tips of velocity
vectors corresponds to a speed $0.5 \, \cs$ in the top panels, 
to $1.0\,\cs$ in the bottom left panel, and to $3.0\,\cs$ in the
bottom right panel.
Spatial coordinates are normalized to $2 \pi H$, the 
wavelength of maximum growth rate in the limit of no magnetic field 
and external pressure.
}
\label{basufig1}
\end{figure}

Figure \ref{basufig1} shows a comparison of column density structure and velocity
vectors for four different models (from Basu, Ciolek, \& Wurster 2008;
Basu et al. 2008). Each image is shown at the time of
runaway collapse of the first supercritical dense core that forms in the
simulation, with a column density enhancement of 
a factor of 10. The top two images are for subcritical (left)
and supercritical (right) clouds evolving from linear initial 
perturbations. The bottom two images are for subcritical (left)
and supercritical (right) clouds evolving from highly nonlinear (turbulent)
initial conditions. Each image represents a physically distinct path
to core/star formation.
Velocity vectors are overlaid, although the normalizations may differ
(see caption). 
The subcritical models have $\mu_0=0.5$ (i.e., the mass-to-flux ratio 
is half the critical value)
and the supercritical models have $\mu_0=2$. 
Amongst the gravitational fragmentation models (linear initial perturbations),
there is extended and mildly supersonic motions only in the 
supercritical model, whereas the subcritical model has a maximum speed
of only $\sim 0.4\cs$. 
For the subcritical gravitational fragmentation model the runaway 
collapse to the first core occurs at a time 7.4 Myr, whereas for the
supercritical model it occurs at 0.83 Myr, assuming a background 
column density $10^{22}$ cm$^{-2}$ and temperature 10 K. 
Fragmentation of a transcritical ($\mu_0 \simeq 1$) cloud is qualitatively
unique in that the fragmentation spacing is much larger than the typical  
value $\simeq 2 \pi H$ that is valid for both highly supercritical and
highly subcritical clouds (Ciolek \& Basu 2006). The timescale of 
transcritical fragmentation is intermediate to the subcritical and 
supercritical cases, but closer to the former.
Here, we focus primarily on the distinction between decidedly subcritical and
supercritical clouds.
The bottom panels of Fig. \ref{basufig1} show the corresponding $\mu_0=0.5$ 
(left) and $\mu_0=2$ (right) models, now with turbulent initial perturbations 
with rms amplitude $v_a = 2\cs$ in each of $v_x$ and $v_y$.
The power spectrum is such that $v_k^2 \propto k^{-4}$, so most of the energy
is in the largest scale modes: the result is immediate large-scale compressive
motions in the cloud. 
The $\mu_0=0.5$ model undergoes rapid ambipolar diffusion during the 
first compression, but still rebounds due to stored magnetic energy.
It subsequently oscillates several times before continuing ambipolar
diffusion causes runaway collapse in the highest density region.
The total time for this process is 1.2 Myr, similar to that of supercritical
gravitational fragmentation. In contrast, the $\mu_0=2$ model
goes into prompt collapse during the first compression, on a timescale
of merely $4.0 \times 10^4$ yr $= 0.04$ Myr.  
Systematic supersonic motions exist in both turbulent models, but the
relative infall is generally subsonic in the subcritical model.
A comparison of the two bottom panel images shows that the supercritical
model is extremely filamentary, while the subcritical model is much less so
since it has had a chance to rebound from the initial compression.
These turbulent simulations are done in the thin-disk approximation
and are consistent with the earlier results of Li \& Nakamura (2004)
and Nakamura \& Li (2005), also done using the same approximation.
Recent fully three-dimensional simulations (Kudoh et al. 2007;
Kudoh \& Basu 2008) also confirm the results.
A summary of main outcomes of the four different modes of core/star 
formation is given in Table \ref{basutable1}.

\begin{figure}
\begin{center}
\includegraphics[width=8cm]{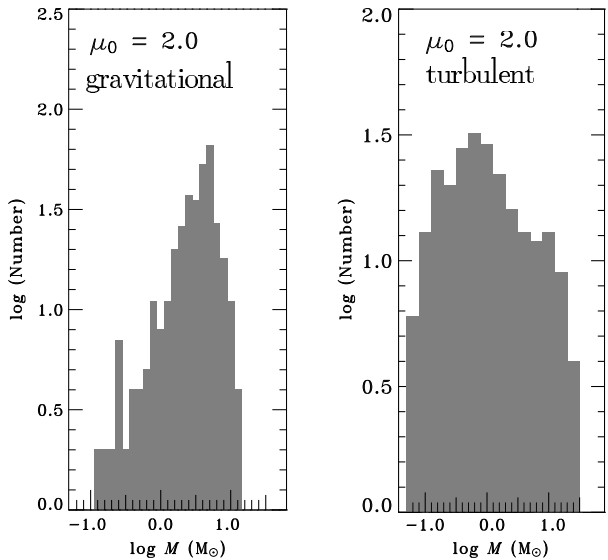} 
\end{center}
\caption{Histograms of masses contained within regions with column density
enhancement above a factor of two, measured at the end point of simulations
with $\mu_0 = 2.0$. The left panel corresponds to gravitational 
fragmentation and the right panel corresponds to turbulent fragmentation.
See text for details.
Each figure is obtained from the compilation of results of a large number of
simulations. The bin width is 0.1.}
\label{basufig2}
\end{figure}

The results described above are used to generate statistics of core 
masses, sizes, shapes, etc. Each model with a unique set of parameters 
is run $\sim 100$ times in order to generate a meaningful measure of 
the various outcomes arising from different realizations of the initial 
random perturbations. Thresholding techniques are used to identify cores -
for details of the technique, see Basu, Ciolek, \& Wurster (2008).
Figure \ref{basufig2} shows a comparison of the initial core mass function (CMF) for 
supercritical models with $\mu_0=2$ but without and with turbulent initial
perturbations, respectively. We emphasize these are {\it initial} CMF's because they 
reflect the status of cores at the time that the first supercritical 
dense core undergoes runaway collapse.
The clear distinction to be made between the two models is
that the linear initial perturbation case (left panel) has an extremely
sharp peak (consistent with a preferred mass scale for fragmentation in
the linear theory), although there is a broader tail at 
the low mass end due to some cores that are very young and just 
emerging above the threshold.
In contrast, the turbulent initial condition case develops
a broad tail of high mass cores, due to much of the turbulent power
being on large scales.
However, the high mass cores often have quite disturbed structures 
(see Fig. \ref{basufig1} lower panels) and it is not clear that they would 
subsequently collapse monolithically.

The narrow initial CMF seems to be a problem for purely gravitational 
fragmentation models. However, a broader distribution may yet be possible
if the newly formed cores continue to accrete from their
environment. If the accretion from the environment is nonuniform from core to core,
an initially narrow and/or lognormal CMF may become skewed so that a high mass
tail develops. For example, Basu \& Jones (2004) have generalized an 
earlier result of Myers (2000) and shown that an 
initially lognormal CMF and an exponential distribution of subsequent 
accretion times from the cloud results in a CMF that is like a lognormal
at low masses but has a power-law tail at high masses.
The model formally requires that the accretion rate onto a core is linearly
proportional to the instantaneous core mass. Additionally, Basu \& Jones (2004) 
show that a different accretion law can also lead to a broadened near-power-law 
tail.

\subsection{Observational diagnostics}
\label{observations} 

In principle, it should be possible to discriminate between these various core formation scenarios 
based on detailed observational studies of the 
characteristics of prestellar cores and young protostars.

\begin{figure}
\begin{center}
\includegraphics[width=7.5cm, angle=270]{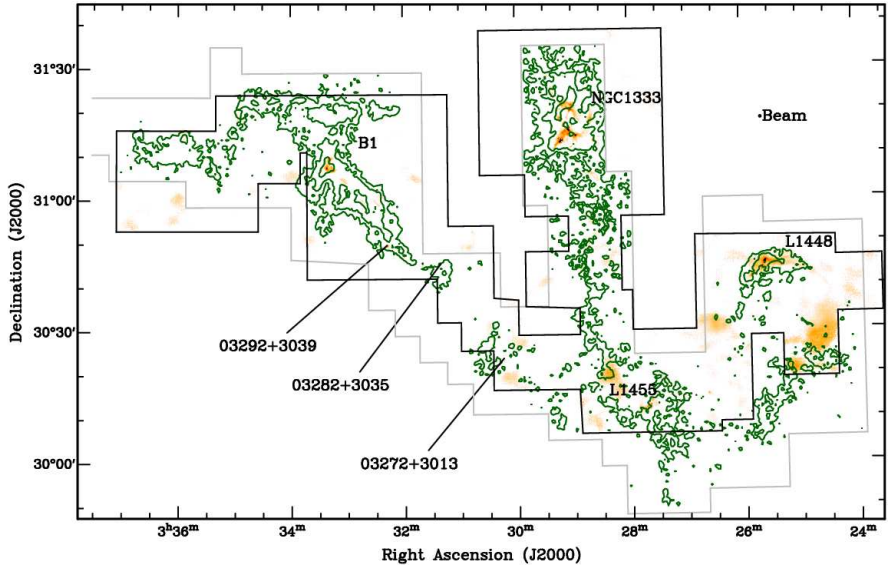}
\end{center}
\caption{SCUBA 850~$\mu$m dust continuum map (colorscale) of the Western part of the Perseus cloud 
complex. Contours of C$^{18}$O(1-0) integrated intensity are overlaid. The black boundary indicates the 
total area mapped with SCUBA. Note that dust continuum cores are detected only localized sub-regions, 
such as NGC~1333, within the complex. (From Hatchell et al. 2005) }
\label{scuba_perseus_fig}
\end{figure}

\noindent
\subsubsection{Core formation efficiency and spatial distribution from surveys}
\label{surveys}

A number of very large submillimeter continuum surveys of nearby cloud complexes have been completed 
recently which provide the spatial distribution of cores within these complexes and set constraints on the 
efficiency of the core formation process (e.g. Johnstone et al. 2004, Hatchell et al. 2005, Enoch et al. 2006, 
Motte et al. 2007).
As an illustration, Figure~\ref{scuba_perseus_fig} shows part of the 3~deg$^2$ SCUBA 850~$\mu$m survey 
of the Perseus cloud complex by Hatchell et al. (2005).
Such extensive surveys show that prestellar cores and Class~0 protostars are found in localized 
sub-regions within molecular clouds which occupy only a very small fraction of their volume. 
These localized active sub-regions often correspond to cluster-forming clumps associated with embedded 
near-IR clusters. 
On this basis, it has been suggested that there may be a threshold in background column density (or 
equivalently visual extinction at $A_V \sim 5-10$) for core formation (Onishi et al. 1998, Johnstone et al. 2004).
Observationally, however, establishing the presence (or absence) of such a threshold is difficult since there 
are also detection thresholds. According to Hatchell et al. (2005), there is no real threshold but the probability 
of forming a prestellar core is a steeply rising function of background column density.
In any case, the results of these wide-field surveys for cores clearly demonstrate the global inefficiency of the 
core formation process: the fraction of cloud mass observed in the form of prestellar cores is very low 
($\simlt 1- 20\% $  --  Hatchell et al. 2005, Nutter et al. 2006). Furthermore, there is some evidence that 
core/star formation in the observed cluster-forming clumps has been induced by external triggers 
(Nutter et al. 2006, Kirk et al. 2006).  These results are broadly consistent with the view that the low-density envelopes  
of molecular clouds are supported against collapse by magnetic fields, as in the classical ambipolar diffusion 
picture (Mouschovias 1987, Shu et al. 1987, McKee 1989).

\noindent
\subsubsection{Core lifetimes}
\label{lifetimes}

\begin{figure}
\begin{center}
\includegraphics[width=10cm]{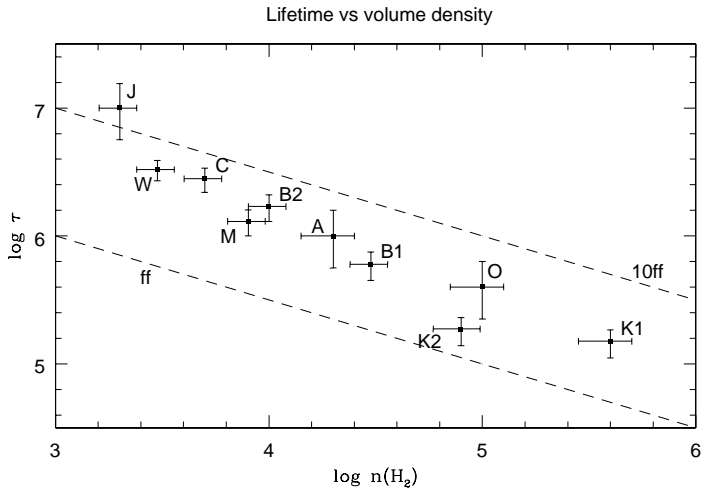}
\end{center}
\caption{
Plot of inferred core lifetime against mean volume density for various samples of starless and 
prestellar cores. The two dashed lines correspond to one and ten free-fall times, respectively.
(From Jessop \& Ward-Thompson 2000 and Kirk et al. 2005.)
}
\label{lifetime_fig}
\end{figure}

In the turbulent paradigm, cloud cores are always dynamically evolving (e.g. Dib et al. 2007) and 
survive for at most a few free-fall times.\\
Observationally, a rough estimate of the lifetime of starless cores  can be obtained from the number ratio 
of cores with and without embedded YSOs in a given core sample. 
Using this technique, Lee \& Myers  (1999) found that the typical lifetime of starless 
cores with average volume density $\sim 10^4\ \rm{cm}^{-3}$ was $\sim 1-1.5 \times 10^6$~yr. 
Furthermore, by considering several samples of isolated cores spanning a range of core densities, 
Jessop \& Ward-Thompson (2000) established that the typical core lifetime decreased as the mean volume 
density in the core sample increased (Fig.~\ref{lifetime_fig} -- see also Kirk et al. 2005).
As can be seen in Fig.~\ref{lifetime_fig}, all of the observed lifetimes lie between one free-fall time, which 
is the timescale expected in free-fall collapse, and ten free-fall times, the timescale expected for highly subcritical 
cores undergoing ambipolar diffusion. The observed timescales are typically longer than the free-fall time by a
factor of $\sim$2--5 in the density range of 10$^4$--10$^5$ cm$^{-3}$. This suggests that starless cores cannot all be 
rapidly evolving, at variance with purely hydrodynamic scenarios of core formation 
(e.g. Ballesteros-Paredes et  al. 2003), but in agreement with numerical simulations of moderately 
supercritical, turbulent molecular clouds (Galv\'an-Madrid et al. 2007).
The observed timescales also appear to be too short for all of the cores to form from a highly 
subcritical state. These statistical estimates of core lifetimes are however quite uncertain  
since they assume that all of the observed cores follow the same evolutionary path and that the core/star formation 
rate is constant.

\noindent
\subsubsection{Magnetic field measurements}
\label{bfield}

\begin{figure}
\begin{center}
\includegraphics[width=9cm]{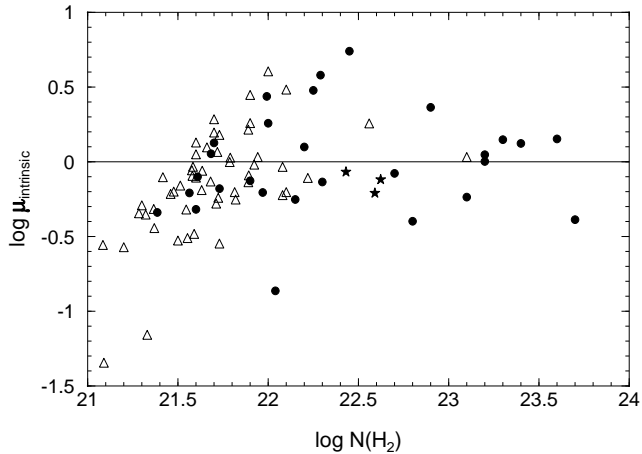} 
\end{center}
\caption{Plot of observed mass-to-magnetic flux ratio,  in units of the critical value and divided by 3 to correct for projection
bias, against cloud/core column density. Dots are for Zeeman detections of B$_\parallel$ (above 3$\sigma$); stars are for 
Chandraskehar-Fermi estimates of B$_\perp$; open triangles are lower limits 
(corresponding to observed upper limits to the field strength). (From Heiles \& Crutcher 2005 -- see also 
Crutcher 2004 for details.)}
\label{magnetic_fig}
\end{figure}

In principle, observations of magnetic fields can provide a strong discriminator
between the two extreme paradigms of the core/star formation process. 
Two main observational techniques have been used to estimate the magnetic field 
strength in cloud cores.
First, the Zeeman effect in, e.g., the 18-cm lines of OH,  gives  
a direct measurement of the line-of-sight component of the magnetic field, but relatively 
few positive detections have been obtained. 
For existing detections, a very good correlation is found between the magnetic field strength $B_{los}$ 
and $n^{0.5} \, \sigma_v$, where $n$ is the gas density and $\sigma_v$ is the line-of-sight 
velocity dispersion observed in optically thin molecular line tracers (Crutcher 1999, Basu 2000).
This shows that magnetic fields play a dynamically important role during the contraction
of at least some cloud cores and is consistent with cloud turbulence being MHD in character.
Second, maps of linearly polarized dust emission at submillimeter wavelengths, combined 
with the Chandrasekhar and Fermi method, provide an indirect estimate of the 
plane-of-the-sky component of the magnetic field (e.g. Ward-Thompson et al.  2000, 
Crutcher et al. 2004). 
The available B-field measurements based on either the Zeeman technique or the 
Chandrasekhar and Fermi method are summarized in Fig.~\ref{magnetic_fig} which plots the 
inferred mass-to-magnetic flux ratio corrected for projection effects against the column density of 
each core (see Crutcher 2004 and Heiles \& Crutcher 2005). It can be seen that all cloud cores 
are scattered around the critical mass-to-flux ratio in this plot, suggesting that, on average,  cores 
are close to magnetically critical. Furthermore, there is some hint in Fig.~\ref{magnetic_fig} that the 
mass-to-flux ratio may be systematically subcritical at column densities lower than 
$\sim 3 \times 10^{21}\ \rm{cm}^{-3}$. This tentative trend is weak, however, as it relies only on the 
locations of Zeeman non-detections in the diagram.
Thus, existing magnetic field measurements do not allow a definite conclusion to be drawn,  
even if they seem to favor pictures of the core formation process in which the magnetic field does play 
an important role.

\noindent
\subsubsection{Radial density structure}

The density profiles of isolated prestellar cores are now fairly well known.
Two methods have been used: (1)~mapping the optically thin 
(sub)millimeter continuum {\it emission} from the cold dust contained 
in the cores, and (2) mapping the same cold core dust in {\it absorption} 
against the background infrared emission (originating from warm cloud dust 
or remote stars). 

Ward-Thompson et al. (1994, 1999) and Andr\'e et al. (1996)  
employed the first approach to probe the structure of prestellar cores
(see also Shirley et al. 2000).
Under the simplifying assumption of spatially uniform dust temperature and
emissivity properties, they concluded that the radial density profiles of
isolated prestellar cores  
were flatter than $\rho(r) \propto r^{-1}$ in their inner regions (for $r \leq R_{flat}$), 
and approached $\rho(r) \propto r^{-2}$ only beyond a typical radius
$R_{flat} \sim $~2500--5000~AU. 

More recently, the use of the {\it absorption} approach, 
both in the mid-IR from space (e.g. Bacmann et al. 2000) and in the near-IR from the ground 
(e.g. Alves et al. 2001), made it possible to confirm and extend the 
(sub)millimeter emission results, essentially independently of any assumption
about the dust temperature distribution. In some cases, such as L1689B, the absorption 
studies indicate that isolated prestellar cores feature sharp edges defining outer radii 
$R_{out} \sim $~0.1~pc (cf. Bacmann et al. 2000). 

The circularly-averaged column density profiles 
can often be fit remarkably well with models of pressure-bounded 
Bonnor-Ebert spheres, as first demonstrated by Alves et al.~(2001) for B68. 
This is also the case of cores detected in the submillimeter continuum such as L1689B 
(e.g. Kirk et al. 2005).
The quality of the fits shows that equilibrium 
Bonnor-Ebert spheroids provide a good, first order model for the structure of
isolated prestellar cores. In detail, however, there are problems with
this model. 
First, the inferred density contrasts (from center to edge) 
are generally larger (i.e., $\simgt $~20--80 -- cf. Bacmann et al. 2000 and Kirk et al. 2005) 
than the maximum contrast of $\sim $~14 for stable Bonnor-Ebert 
spheres.
Second, the effective gas temperature needed in the fits 
is often significantly larger than measured core temperatures 
(e.g. Ward-Thompson et al. 2002, Lai et al. 2003). 
These arguments suggest that prestellar 
cores are either already contracting 
(see Lee, Myers, Tafalla 2001 and  \S ~\ref{velocity} below) 
or experiencing extra support from static or turbulent magnetic fields 
(e.g. Curry \& McKee 2000).
As shown by Bacmann et al. (2000), 
one way to account for large density contrasts and high effective
temperatures is to consider models of cores initially supported 
by a static magnetic field and evolving through ambipolar diffusion 
(e.g. Ciolek \& Mouschovias 1994, Basu \& Mouschovias 1994). 

However, good Bonnor-Ebert fits can often be found for dynamically evolving  
``cores'' produced by turbulent compression (Ballesteros-Paredes et al. 2003, G\'omez et al. 2007).
Thus, the observed density profiles do not provide a very strong diagnostic of proposed
core formation models.

\subsubsection{Velocity structure}
\label{velocity}

\begin{figure}
\begin{center}
\includegraphics[width=12.5cm]{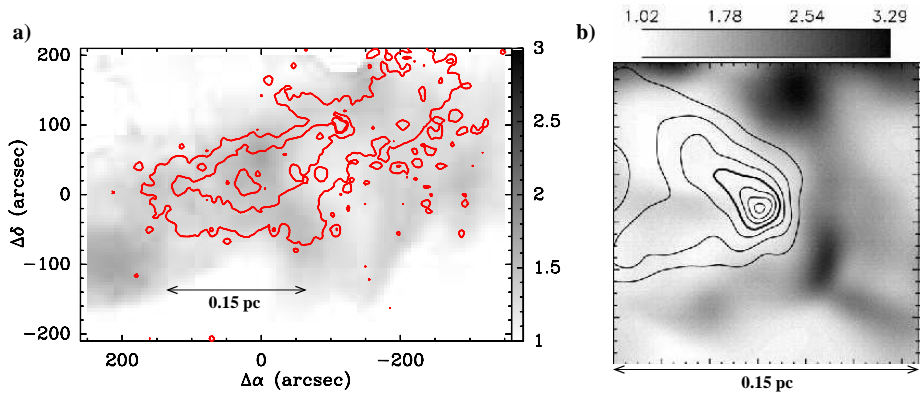}
\end{center}
\caption{\rm{a)} Greyscale image of the line-of-sight velocity dispersion (in units of the isothermal sound speed) derived 
from deep $^{13}$CO(1-0) line observations toward and around the 
prestellar core L1689B (from Andr\'e, Pety, \& Bacmann, in prep.). 
Column density contours from Bacmann et al. (2000) are superimposed and delineate  the core boundaries. 
The maximum observed value of $\sigma_{los}$  is only $\sim 0.34$~km/s or $\sim 1.7\, c_s$
\rm{b)} Greyscale image of the line-of-sight velocity dispersion superimposed on column density contours for 
a model core obtained in SPH numerical simulations of  gravoturbulent fragmentation (from Klessen et al. 2005). 
Note that these simulations produce localized maxima where $\sigma_{los}  \simgt 3\, c_s$ at the shock 
positions.
}
\label{veldis_fig}
\end{figure}

Observing the velocity field within and around dense cores is probably the most effective 
way to discriminate between quasi-static and dynamic core formation scenarios. 
If cores are produced by shocks in large-scale supersonic flows, large-velocity gradients and local maxima of 
the line-of-sight velocity dispersion are expected in the immediate vicinity 
of cores (Ballesteros-Paredes et al. 2003; Klessen et al. 2005). 
A much more quiescent ambient velocity field is expected in the  
magnetically-controlled picture (e.g. Nakamura \& Li 2005 -- see also \S ~\ref{theory} and Fig.~\ref{basufig1}).

Briefly, isolated starless cores are characterized by subsonic levels of internal turbulence 
(e.g. Myers 1983, Goodman et al. 1998, Caselli et al. 2002), small rotational velocity gradients 
(e.g. Goodman et al. 1993, Caselli et al. 2002), and extended, subsonic infall motions 
(e.g. Tafalla et al. 1998, Lee et al. 2001). Moreover, the environment of isolated prestellar cores 
is also extremely quiescent. This has been shown recently through deep mapping of several cores 
in low-density molecular gas tracers such as $^{13}$CO(1--0) and C$^{18}$O(1--0).
Figure~\ref{veldis_fig}a shows a greyscale map of the line-of-sight velocity dispersion $\sigma_{los}$ observed in  
$^{13}$CO(1--0) toward the prestellar core L1689B in the Ophiuchus complex (Andr\'e et al. in prep.). 
It can be seen that $\sigma_{los}$ remains at most transonic ($\sigma_{los} < 2\, c_s$, where $c_s$ is the 
isothermal sound speed) everywhere in a region of more than 0.25~pc in diameter around the column
density peak. Such a quiescent velocity field is at variance with purely hydrodynamic models of gravoturbulent 
fragmentation (e.g. Klessen et al. 2005). While these models successfully produce a fair amount ($\sim 25\% $) 
of cores with subsonic internal velocity dispersions, corresponding to dense, post-shock stagnation points at the  
intersection of converging flows, they also produce highly supersonic maxima of $\sigma_{los}$ in the low column 
density gas around the cores (cf. Fig.~\ref{veldis_fig}b), which are not observed.
Current observations therefore provide strong indirect evidence that the evolution of isolated prestellar cores is 
magnetically controlled. 

The environment of individual cores in cluster-forming regions is known to be more turbulent 
(e.g. Caselli \& Myers 1995), so that hydrodynamic core formation models may be more appropriate in this case.
Indeed, observations of Class~0 objects indicate that protostellar collapse is more dynamic, with 
supersonic infall velocities and large mass accretion rates ($>10\, c_s^3/G$), 
in cluster-forming clumps (e.g. Di Francesco et al. 2001, Belloche et al. 2006). 
Evidence of coherent, supersonic contraction motions over more than 0.5~pc has even been found in some 
protoclusters (e.g. Motte et al. 2005, Peretto et al. 2006).
On small scales, however, the compact ($\sim 0.03$~pc) 
prestellar condensations of the Ophiuchus, Serpens, Perseus, and Orion  
protoclusters are themselves characterized by subsonic levels of internal turbulence 
(Myers 2001, Andr\'e et al. 2007) reminiscent of the thermal cores 
of Taurus. Furthermore, the condensation-to-condensation velocity dispersion measured in these 
cluster-forming regions is small and only subvirial (Andr\'e et al. 2007). This is consistent with 
the view that protoclusters often start their evolution from ``cold'', out-of-equilibrium initial 
conditions (cf. Adams et al. 2006), perhaps as a result of external perturbations (cf. Nutter et al. 2006).
The observed small relative velocity dispersion also implies that collisions between condensations in 
a low-mass protocluster such as L1688 are relatively rare and that, in general, prestellar condensations 
do not have time to interact with one another before evolving into pre-main sequence objects 
(Andr\'e et al. 2007). Furthermore, in such protoclusters, the mass accretion rate expected from competitive, 
Bondi-like accretion of background gas onto the condensations (e.g. Bonnell et al. 2001)
is estimated to be at least a factor $\sim 3$ lower than the mass infall rate resulting from gravitational 
collapse at the Class~0 and Class~I stages (Andr\'e et al. 2007 -- see also Krumholz et al. 2005). 
Therefore, competitive accretion cannot play a dominant role once individual protostellar collapse 
sets in. On the other hand, Bondi-like accretion of unbound gas is more effective before protostellar collapse, 
and may possibly govern the growth of starless, self-gravitating condensations initially produced by 
gravitational fragmentation (cf. \S ~\ref{theory} and Fig.~\ref{basufig2}) 
toward a Salpeter-like IMF mass spectrum (cf. Myers 2000, Basu \& Jones 2004, Clark \& Bonnell 2005).

\section{Collapse and subfragmentation of prestellar cores}
\label{collapse}

\subsection{Core collapse models: Thermodynamics}

The evolution of gravitationally collapsing cores and the formation 
 of protostars are radiation-magnetohydrodynamical processes. 
Although we need to model these processes by solving 
 equations of radiative transfer and magnetohydrodynamics 
 simultaneously in multi-dimensions,  a direct calculation 
 of all the equations remains challenging. 
The most sophisticated theoretical models so far are either 
 non-magnetic radiation hydrodynamical calculations based on the 
 (flux-limited) diffusion approximation  
 or  magnetohydrodynamical calculations with some prescribed equations of 
 states. 

Here, we first explain the thermodynamics 
 of gravitational collapse revealed by 
 dynamical modelling with detailed radiative transfer 
 in {\em spherical} symmetry 
 (Larson 1969;  
  Narita, Nakano, \& Hayashi 1970; 
  Winkler \& Newman 1980; 
  Stahler, Shu, \& Taam 1980; 
  Masunaga, Miyama, \& Inutsuka 1998; 
  Masunaga \& Inutsuka 2000a).
The classical results based on gray-approximation 
 were confirmed by recent work 
 that solved the frequency-dependent RHD equations by the 
 ``Variable Eddington Factor Method'' (Masunaga \& Inutsuka 2000a). 
The latter provides the time evolution of the apparent spectrum 
 of the radiation field (Spectral Energy Distribution -- SED) 
 in addition to the detailed dynamical evolution of 
 the protostar.  

\begin{figure}
\begin{center}
  \includegraphics[width=100mm]{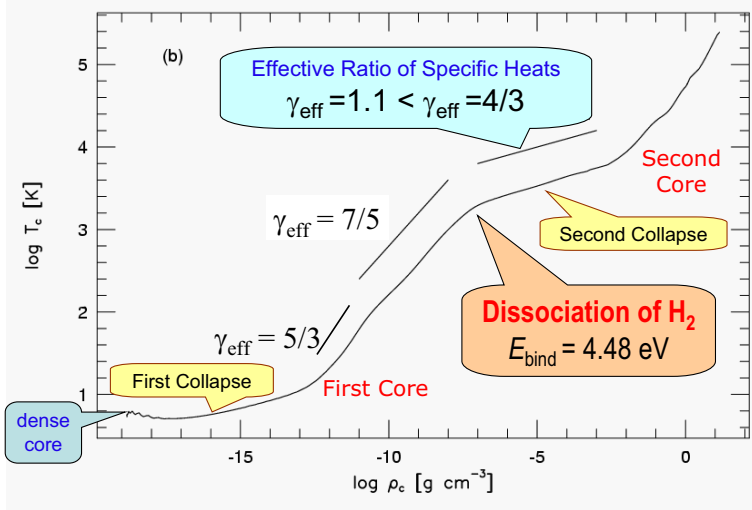}
\end{center}  
  \caption{
Temperature evolution at the center of a gravitationally 
collapsing cloud obtained by Masunaga \& Inutsuka (2000a) in 
their radiation hydrodynamical calculation of protostellar 
collapse in spherical symmetry.
The {\em first collapse} phase corresponds to the formation 
of the {\em first protostellar core} that consists mainly of hydrogen molecules.  
The dissociation of hydrogen molecule triggers the 
{\em second collapse} that eventually produces  
the {\em second core}, i.e., a protostellar object. 
Each of these phases in the temperature evolution is characterized by a distinct value 
of  the effective ratio of specific heats, $\gamma_{\rm eff}$. 
  }
  \label{fig:Tc}
\end{figure} 
 
Figure~\ref{fig:Tc} shows the time evolution of 
the central temperature (as a function of the central density) 
in a collapsing cloud.  
Once compressional heating dominates over radiative cooling, 
the central temperature increases gradually above the low value 
($\sim 10$ K) found in molecular clouds.  
At first, the slope of the 
temperature curve corresponds to a ratio of specific heats $\gamma_{\rm eff} =5/3$: 
$T(\rho)\propto \rho^{2/3}$ for $10\, {\rm K}<T<10^2\, {\rm K}$. 
This apparently monoatomic gas behavior is due to the fact that 
the rotational degree of freedom of molecular hydrogen 
is not excited in this low temperature regime 
($E(J=2-0)/k_{\rm B} = 512$~K). 
When the temperature rises above $\sim 10^2$~ K, 
the slope becomes that of diatomic molecules ($\gamma_{\rm eff} =7/5$). 
In both cases, the effective ratio of specific heats is larger than 
the critical value for gas pressure support against 
self-gravity:  
$\gamma_{\rm eff}~>~\gamma_{\rm crit}~\equiv~4/3$. 
Thus, the collapsing velocity is decelerated and forms 
a shock at the surface of quasi-adiabatic hydrostatic object, 
the ``first core''. 
Its radius is about 1~AU in spherically symmetric calculations 
 (but is larger by an order of magnitude in 2D/3D calculations with 
 rotation).  
It mainly consists of H$_2$. 
The increase in density and temperature inside 
the first core is slow but monotonic. 
When the temperature becomes $>~10^3$~K, 
the dissociation of H$_2$ starts. 
The binding energy of H$_2$ is about 4.5~eV which is much larger 
than the thermal energy per hydrogen molecule 
in this temperature regime. 
Therefore the dissociation of H$_2$ acts as an efficient 
coolant of the gas, 
which reduces the effective ratio of specific heats below the critical value 
($\gamma_{\rm eff}~<~4/3$), 
and triggers the second dynamical 
collapse\footnote{
 Note that the role of H$_2$ dissociation in this second 
 collapse is analogous to that of the photo-disintegration of 
 Fe as the cause of the pre-supernova gravitational collapse.}. 
In this ``second collapse'' phase, the collapsing velocity becomes 
 very large and engulfs the first core.   
As a result, the first core lasts only for $\sim 10^3$~yr. 
In the course of the second collapse, 
 the central density reaches a stellar value ($\rho_\star \sim 1\ \rm{g}\,\rm{cm}^{-3}$), and  
a truly hydrostatic protostellar object forms in the center.

Radiation hydrodynamic (RHD) calculations 
 automatically produce the time evolution of the accretion luminosity and 
 SED of the collapsing object (cf. Masunaga \& Inutsuka 2000a).
The resultant luminosity evolution has a sharp growth at the 
 formation of the first core and a peak around the formation of 
 the second protostellar core in the case of dynamical 
 initial conditions, while it only shows a gradual growth in the case 
 of hydrostatic equilibrium initial conditions.   
This difference in the  time evolution of the accretion luminosity may provide 
 a useful observational diagnostic. 

Molecular emission line profiles of various 
 important species were also calculated in self-consistent 
 dynamical models (Masunaga \& Inutsuka 2000b). 
Recent three dimensional modelling of protostellar radiation 
 hydrodynamics can be found, e.g., in 
 Whitehouse \& Bate (2006), Stamatellos et al. (2007b), 
 and Krumholz et al. (2007). 

Further evolution includes the T-Tauri phase on Hayashi tracks,  
for which the relevant (Kelvin-Helmholtz) timescale is about two orders of magnitude larger than 
 the dynamical timescale ($\sim 10^5$~yr) of protostellar collapse, 
 which is only accessible by steady state calculations 
 (e.g.  Chabrier \& Baraffe 2000). 

\subsubsection{Dynamical roles of the first and second protostellar cores}
The formation of the `short-lived' first core plays 
an important role in the dynamical evolution of 
protostellar collapse, as well as the second core does. 
This is because the associated pressure support helps 
 the formation of a rapidly rotating disk-like structure around 
 the central object in the presence of  non-zero initial angular momentum 
 (Saigo, Matsumoto, \& Hanawa 2000). 

\medskip
\paragraph{{\bf Rotationally-driven fragmentation  into multiple systems:}}

The mechanisms responsible for the formation of multiple systems have been the subject of 
 extensive theoretical work (see, e.g., Goodwin et al. 2007 for a review).
The essence of the results may be summarized as follows. 
First, gravitational sub-fragmentation within a collapsing cloud 
 core is not expected during the early isothermal phase of evolution, 
 unless the mass of the core is much larger than the initial 
 Jeans mass.  
This result has been found in numerical simulations 
 and is supported by semi-analytic calculations 
 in the case of an initially uniform cloud with solid rotation  
 (Tsuribe \& Inutsuka 1999a,b). 
Initial central concentration in the density profile makes 
core fragmentation even more difficult.  
However, if the evolution is followed to the higher density regime 
where the gas becomes adiabatic, a disk-like structure forms 
which allows another mode of binary formation to develop, 
 i.e., disk fragmentation around the central protostar.   
For example, calculations based on a piecewise 
 polytropic equation of state show that 
 the central portion of a collapsing core becomes adiabatic
 and forms a disc-like structure around the central object,
 which subsequently fragments into ``satellite'' objects 
 (Matsumoto \& Hanawa 2003).   
The study of  realistic radiative transfer effects 
 on these modes of binary fragmentation remains an important task  for future work. 
The above phenomena correspond to the fragmentation of/around 
 the first core, which might account for the formation of binaries 
 with separations $\simgt 1$AU. 
Obviously formation of binaries with even shorter separations 
is also expected in the disk around the second protostellar core.  
These multiple epochs of core fragmentation may result in 
multiple peaks in the separation distribution of binary stars
as proposed by Machida et al. (2007c). 

\noindent
The effects of initial turbulence on binary fragmentation have also
been studied extensively (see, e.g., Goodwin et al. (2007). 
Purely hydrodynamic SPH simulations of rotating cloud core collapse show 
that a very low level of initial core turbulence 
(e.g. $E_{\rm turb}/E_{\rm grav} \sim $~5\%)  leads to the formation of a 
multiple system (Goodwin et al. 2004; Hennebelle et al. 2004). 
In such hydrodynamic simulations, fragmentation is driven by a combination of 
rotation/turbulence and occurs in 
large ($\simgt 100$~AU) disk-like structures or ``circumstellar accretion regions'' 
(CARs -- cf. Goodwin et al. 2007). 
These CARs are highly susceptible 
to spiral instabilities which {\it always} fragment them into small-$N$ {\it multiple systems  
with $N > 2$} and typically $N \sim $~3-4 within a radius $\sim 150$~AU 
(Goodwin et al. 2004).  
However, recent MHD simulations of {\it magnetized} core collapse (Price \& Bate 2007, 
Hennebelle \& Teyssier 2008) show that 
the presence of an even moderate magnetic field strongly modifies 
angular momentum transport during collapse and at least partly suppresses core 
fragmentation. Therefore, it is unclear at the present time whether the collapse of an individual 
prestellar core typically produces one, two, or more stars. 

\medskip
\paragraph{{\bf Implications for the CMF-IMF connection:}}

If each prestellar core is the progenitor of $N \sim 2-4$ stars,  
then the IMF of individual stars 
cannot be the direct product of the CMF but results instead from the convolution of the 
CMF with the typical distribution of object masses produced by binary fragmentation for 
one mass of core (cf. Delgado-Donate et al. 2003, Goodwin et al. 2008). 
For realistic binary fragmentation scenarios, 
the IMF will still follow the CMF at the high-mass end (because the majority of each core's mass 
can still end up in one stellar component), but may differ substantially from the CMF 
at the low-mass end. Such a picture is consistent with present determinations of the 
CMF (see \S ~\ref{sec:surveys}).
However, it is presently unclear  whether the collapse of an individual 
prestellar core typically produces one, two, or more stars. 
Accordingly, the origin of the low-mass end ($\simlt 0.1\, M_\odot $) of the IMF is highly uncertain. 
Observationally, this is an area where the future large millimeter interferometer ALMA will yield key progress.

\medskip
\paragraph{{\bf Driving MHD outflows:}}

Another effect of the formation of the first core 
occurs in the MHD evolution of 
 the self-gravitating collapsing core.  
A magnetically supercritical core whose rotation axis is 
 parallel to the mean direction of the magnetic field lines 
 leads to self-similar collapse  
 as long as the equation of state is isothermal 
 (Basu \& Mouschovias 1994).
Once the first core is formed, 
 a rapidly rotating disk-like structure develops 
 owing to the change in 
 the effective equation of state,  
 and its rapid rotation winds up the field lines
 creating a significant amount of toroidal 
 magnetic field.  
This enhanced toroidal magnetic field produces 
 a bipolar outflow driven by magnetic pressure 
 (Tomisaka 2002).  
Thus, the formation of the first hydrostatic core 
 plays a critical role in launching the protostellar outflow. 
A similar process happens again around the second core, 
with higher ejection velocities reminiscent 
of the observed optical jets and high-velocity neutral winds  
 (Machida et al. 2006, 2007a, 2007b). 

The observational detection of the first core would not only 
confirm the predictions of RHD  
modelling, but would also set strong constraints on 
MHD models of protostellar outflows  
as described in the next section. 

\newpage
%----------------------------------------------------------------------%
\noindent
\subsection{Resistive MHD effects and onset of outflows}
%----------------------------------------------------------------------%
%----------------------------------------------------------------------%
\subsubsection{MHD modelling with resistivity}
%----------------------------------------------------------------------%
The full modelling of magnetohydrodynamical processes in star formation 
must include non-ideal MHD effects 
 (e.g. Nakano et al. 2002, Tassis \& Mouschovias 2007). 
Ambipolar diffusion is important at early times during 
the low-density core formation phase, but 
Ohmic dissipation is more important at later times  
in the high-density collapse phase. 
The Hall current term can also be  important 
 in an intermediate regime (Wardle 2004). 
To account for the dissipation of magnetic fields during the 
 formation of protostars, Machida et al. (2006, 2007a, 2007b) used 
 the resistive MHD equation with prescribed resistivity 
 in their three-dimensional nested grid code 
 simulations. 
Their basic equations are as follows: 
\begin{eqnarray} 
& \dfrac{\partial \rho}{\partial t}  + \nabla \cdot (\rho \vect{v}) = 0, & \\
& \rho \dfrac{\partial \vect{v}}{\partial t} 
    + \rho(\vect{v} \cdot \nabla)\vect{v} =
    - \nabla P - \dfrac{1}{4 \pi} \vect{B} \times (\nabla \times \vect{B})
    - \rho \nabla \phi, & 
\label{eq:eom} \\ 
& \dfrac{\partial \vect{B}}{\partial t} = 
   \nabla \times (\vect{v} \times \vect{B}) + \eta \nabla^2 \vect{B}, & 
\label{eq:reg}\\
& \nabla^2 \phi = 4 \pi G \rho, &
\end{eqnarray}
 where $\rho$, $\vect{v}$, $P$, $\vect{B} $, $\eta$, and $\phi$ denote the density, 
velocity, pressure, magnetic flux density, resistivity, and gravitational potential, respectively. 
Machida et al. estimated the resistivity $\eta$ in equation~(\ref{eq:reg})
according to Nakano et al. (2002) and assumed that 
 $\eta$ was a function of density and temperature. 
They further assumed a barotropic equation of state to mimic the 
 temperature evolution shown in Fig.~\ref{fig:Tc}. Hence 
$\eta$ could be expressed as a function of density only:
$ \eta = \ceta \, \eta_0(\rho), $ 
 where $\eta_0(\rho)$ is a function of the central density. 
The initial conditions adopted by Machida et al. correspond to 
a spherical cloud with a critical Bonnor-Ebert density profile having 
a central (number) density 
 $\rho_{c,0} = 3.841 \times 10^{-20} \, 
 \rm{g} \, \cmci$ ($n_{c,0} = 10^4\cmci$). 
In this case the critical Bonnor--Ebert sphere radius, 
$R_c = 6.45\, c_s [4\pi G \rho_{\rm BE}(0)]^{-1/2}$,  
 corresponds to $ R_c = 4.58 \times 10^4$\,AU. 
The total mass inside the critical radius was $M_0=7.6\,\msun$. 
Initially, the cloud was in solid body rotation around the $z$-axis (at a rate $\Omega_0$) 
and had a uniform magnetic field ($B_{\rm init} =17 $$\mu$G) parallel to 
 the $z$-axis (or rotation axis).
To promote contraction, the density was increased by 70\% 
starting from the critical Bonnor-Ebert sphere.
 
The various models investigated by Machida et al. can be characterized by a single 
non-dimensional parameter $\omega$, related to the cloud's initial rotation rate, 
and defined using the central density $\rho_0$ as
$% \begin{equation}
 \omega = \Omega_0/(4 \pi\,  G \, \rho_0  )^{1/2}. 
$% \end{equation}
 The parameter $\omega$, the initial magnetic field strength, 
 the initial angular velocity $\Omega_0$, 
 the ratio of the thermal ($\alpha_0$) and rotational ($\beta_0$)  
 energies to the gravitational energy,
 the final magnetic field strength at the center, 
 and the rotation period of the protostar at the final snapshot of 
 the calculation  
 are summarized in 
 Table~\ref{table:init}, where SR, MR, and RR stand for (initially) slow, medium, and 
 rapid rotator, respectively.
 
 %
%%%%%%%%%%%%%
%%% Table %%%
%%%%%%%%%%%%%
\begin{table}[ht!]   %% Table 1 table:init
\caption{ Model Parameters and Results~$\dag$}
\begin{center}
\begin{tabular}{ccccccccccccccccc|cc}
\hline
%% \multicolumn{2}{c}
 Model  & $\omega$ &  $B_{\rm init}$ {\scriptsize [$\mu$G]} & $\Omega_0$ {\scriptsize [s$^{-1}$]}  &  $\alpha_0$ &$\beta_0$ & $B_{\rm fin}$ {\scriptsize  (kG)} & $P_{\rm fin}$ {\scriptsize (day)} \\
\hline
SR  & $0.003$  & 17& $7.0 \times 10^{-16}$ & 0.5& $3\times10^{-5}$ &  2.18 & 3.0\\
MR  & $0.03 $  & 17& $7.0 \times 10^{-15}$ & 0.5& $3\times10^{-3}$ &  0.40 & 2.1\\
RR  & $0.3  $  & 17& $7.0 \times 10^{-14}$ & 0.5& $3\times10^{-1}$ &  --- & --- \\
\hline
\end{tabular}
 \begin{list}{}{}
 \footnotesize
  \item[]{ $\dag$  
%  \footnote{
 Representing the thermal, rotational, 
 and gravitational energies as $U$, $K$, 
 and $W$, 
 the relative factors against the gravitational energy are 
 defined as $\alpha_0 = U/|W|$, and $\beta_0 = K/|W|$. 
% } 
}
  \end{list}
  \normalsize  
\end{center}
\label{table:init}
\end{table}

%
%----------------------------------------------------------------------%
\subsubsection{Protostellar outflows \& jets}
\begin{figure}[hb!] 
\begin{center}
\includegraphics[width=75mm]{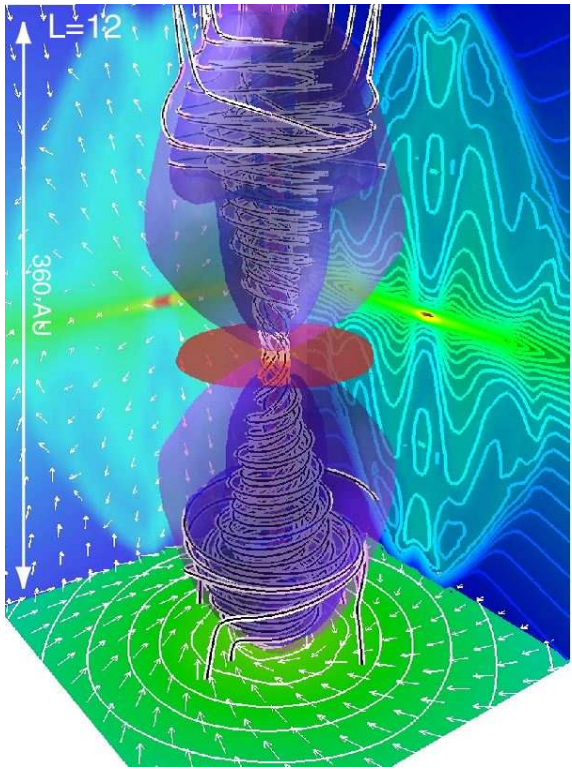}
\caption{
Bird's-eye view of model RR ($l=12$). 
The structure of high-density region ($n > 10^{12} \cmci$; 
iso-density surface), and magnetic field lines 
(black-and-white streamlines) are plotted.
The structure of the outflow is shown by the 
iso-velocity surface inside which the gas is 
outflowing from the center.  
The density contours 
and velocity vectors (thin arrows) on the mid-plane of 
$x=0$, $y=0$, and $z=$0 are, respectively, 
projected in each wall surface.
}
\label{fig:RR}
\end{center}
\end{figure}

%----------------------------------------------------------------------%
As a result of their calculations, Machida et al. found that 
two distinct flows (low- and high-velocity flows) are 
driven by the first and second cores. 
They proposed that the low-velocity flow from the first core 
corresponds to observed molecular outflows, 
while the high-velocity flow from the protostar corresponds to 
observed optical jets.
As an illustration of their simulations, 
a snapshot of Model RR is shown in Fig.~\ref{fig:RR}.

The results of Machida et al. show that the flow driven by the first core 
has  a slow speed and a wide opening angle, while the flow driven by the protostar 
has a high speed and a well-collimated structure. 
The flow speed roughly corresponds to the escape speed of the driving object. 
The difference in the depth of the gravitational potential between 
the first and the second core therefore causes the difference in flow speed.
Typically, observed molecular outflows and optical jets have 
speeds  $v_{\rm out,obs} \simeq 30\kmps$ and $v_{\rm jet,obs}\simeq 100\kmps$, respectively.
At the end of the calculations, 
 the low- and high-velocity flows of Machida et al. only have speeds 
 $v_{\rm LVF} \simeq 3\kmps$ and $v_{\rm HVF}\simeq 30\kmps$, 
 respectively.
However, the first and second cores only have masses 
 $M_{\rm first\,core} = 0.01\, \msun$ and 
 $M_{\rm second\, core} \simeq 10^{-3}\, \msun$, respectively, 
 at the end of the calculations. Each core grows in mass by 
at least 1--2 orders  of magnitude in the subsequent gas accretion phase. 
Since the escape speed increases as the square root of the mass of 
the central object at a fixed radius, 
the speeds of the low- and high-velocity flows may increase 
by a factor of $\sim 10$, and reach  $v_{\rm LVF} \simeq 30\kmps$ and 
$v_{\rm HVF}=300\kmps$, respectively, 
which correspond to typical observed values.

\begin{figure}
\begin{center}
  \includegraphics[width=90mm]{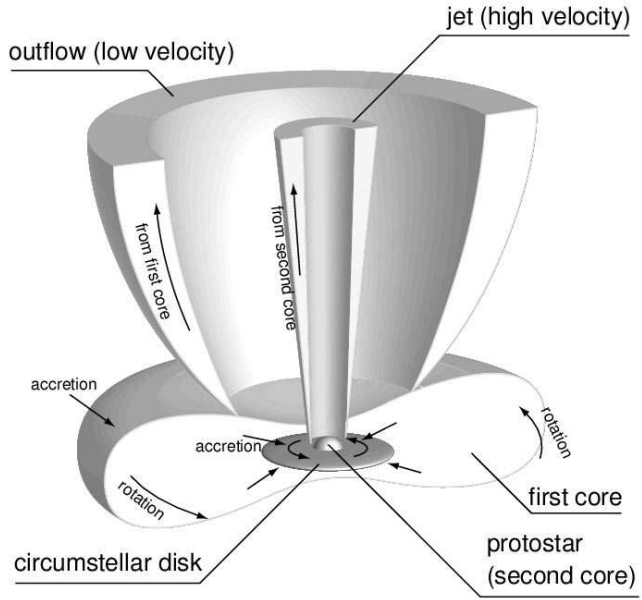}
\end{center}  
  \caption{
 Schematic picture proposed by Machida et al. (2007b) for 
  the jet and outflow driven from the protostar and the fist core, 
  respectively.
    }
    \label{fig:schematic}
\end{figure}

The difference in collimation between the two flows 
 is caused both by different configurations of the magnetic field lines 
 around the driving object and different driving mechanisms 
 (see also Banerjee \& Pudritz 2006 and Hennebelle \& Fromang 2008). 
The magnetic field lines around the first core have an hourglass 
 configuration because they converge to the cloud center 
 as the cloud collapses, and Ohmic dissipation is ineffective 
 before first core formation. Furthermore, 
the flow emerging near the first core is mainly a disk wind
 driven by the magnetocentrifugal wind mechanism.
The centrifugal force is dominant in the low-velocity flow, 
 whereas near the protostar, 
 the magnetic field lines are straight, 
 and the magnetic pressure gradient mechanism is more effective 
 in driving the high-velocity flow. 
The magnetic field lines are stretched by the magnetic tension force 
 near the protostar because the magnetic field is decoupled 
 from the neutral gas.
However, the magnetic field lines are strongly twisted 
 in the region in close proximity to the protostar, 
 where the magnetic field is coupled with the neutral gas again.
Thus, the strong toroidal field generated around the protostar 
 can drive a high-velocity flow, 
 which is guided by the straight configuration of the magnetic field.

Figure~\ref{fig:schematic} summarizes 
the main features of the outflows and jets modelled by 
Machida et al. (2007b). 
Note that the calculations of Machida et al. cover only 
the very early phase of protostar evolution.
Further longer-term calculations are needed to better understand 
the correspondence between the models and observed protostellar outflows.

%----------------------------------------------------------------------%
\subsubsection{Effects of outflows/jets on star-forming cores/clouds}
%----------------------------------------------------------------------%

In principle, outflows and jets from new-born stars may have strong  
dynamical effects on their environments. 
It has been proposed that the violent impact of fast outflows may result 
in the destruction of the cores from which the driving protostars are born 
(Nakano et al. 1995, Matzner \& McKee 2000), 
or even in the dispersal of the parental molecular clouds. 
In particular, the local star formation efficiency at the level of individual cloud cores 
is most likely controlled by the effects of protostellar outflows:  The model calculations of 
Matzner \& McKee (2000) give $\epsilon_{\rm core} \equiv M_\star /M_{\rm core} \sim 25-75\%$ 
for various degrees of core flattening and magnetization.
On a more global level, protostellar outflows may also significantly reduce the fraction of 
total molecular cloud mass going into stars. 
Thus, this process provides an interesting possibility for explaining 
the observed low star formation efficiency in the Galaxy. 

Li \& Nakamura (2006) proposed that protostellar outflows 
could also sustain a high level of turbulence in cluster-forming 
clouds, but Banerjee et al. (2007) argued that this was actually difficult. 
Obviously, a quantitative examination of all these processes 
will require appropriate dynamical modelling of protostellar outflows,
which remains to be done both theoretically and observationally. 

\section{Conclusions: Proposed view for the star formation process}
\label{conclusions}

The prestellar core mass function (CMF) 
appears to be consistent with the stellar IMF between $\sim 0.1\, M_\odot$
and $\sim 5\, M_\odot$, although large uncertainties remain especially at 
the low- and high-mass ends (cf. \S ~\ref{sec:surveys}). Small internal and relative motions 
are measured for these prestellar cores, implying that they are much 
less turbulent than their parent cloud and generally do not have time to interact 
before collapsing to (proto)stars (cf. \S ~\ref{velocity}).  
These results strongly support scenarios according to which the IMF is largely determined 
at the prestellar stage.

None of the extreme scenarios proposed for the formation of prestellar cores can explain 
all observations. Pure ambipolar diffusion 
is too slow to be the main core formation mechanism, for typical levels
of cloud ionization.
Purely hydrodynamic pictures have 
trouble accounting for the inefficiency of core formation and the detailed velocity structure 
of individual cores. A mixed scenario (cf. Nakamura \& Li 2005, Basu et al. 2008) may be 
the solution: supersonic MHD turbulence in a molecular cloud close to magnetic criticality 
generates seed cores, which grow in mass until they become gravitationally unstable and 
collapse in a magnetically-controlled fashion while decoupling from their turbulent environment.
Such a mixed picture has  several advantages. 
On the one hand, relatively strong magnetic fields prevent global collapse and lead 
to inefficient core/star formation on large (GMC) scales. 
On the other hand, the rate of ambipolar diffusion is enhanced by turbulence in 
shock-compressed regions (e.g. Zweibel 2002, Fatuzzo \& Adams 2002, Nakamura \& Li 2005).
The difference between the clustered and the isolated or distributed mode of star formation
may result from local variations around the critical mass-to-flux ratio and/or from the presence 
or absence of external triggers.

Cluster-forming clumps such as L1688 in Ophiuchus or NGC2264-C in Monoceros are likely 
(slightly) magnetically supercritical. There is some evidence that these clumps are in a state of 
global collapse induced by large-scale external triggers (e.g. Peretto et al. 2006, Nutter et al. 2006). 
In this case, the local star formation efficiency within each condensation is high ($\simgt 50\% $ -- Motte et al. 1998) 
and a promising core formation mechanism  is purely gravitational, Jeans-like 
fragmentation of compressed cloud layers (cf. Palous 2007, Peretto et al. 2007), possibly 
followed by subsequent core growth (e.g. Basu \& Jones 2004).

Regions of more distributed and less efficient star formation, such as Taurus or 
the Pipe Nebula, are likely slightly subcritical or nearly critical. 
The size of individual cores is then larger (e.g. Motte et al. 1998), 
the local star formation efficiency is lower ($\sim$~15\%--30\% -- 
Onishi et al. 2002, Alves et al. 2007), and the feedback from protostellar
outflows may be more important in limiting accretion and defining stellar masses 
(e.g. Shu et al. 2004).
 
To fully understand how stars form and how the IMF comes about, a better knowledge 
of the presently poorly known initial conditions for molecular cloud formation is required, 
a subject of growing interest (see, e.g., Hennebelle et al. in this volume).
It is also crucial to further investigate the processes by which prestellar cores 
form, evolve, and eventually collapse and fragment into multiple protostellar systems. 
With present submillimeter instrumentation, observational studies are 
limited by small-number statistics and restricted to the nearest regions.
The advent of major new facilities in the coming years 
should yield several breakthroughs 
in this field. With an angular resolution at 75--300~$\mu$m 
comparable to, or better than, the largest ground-based millimeter-wave  
radiotelescopes, $Herschel$, 
the Far InfraRed and Submillimeter Telescope to be 
launched by ESA in 2008 (cf. Pilbratt 2005), 
will make possible complete surveys for prestellar cores 
down to the proto-brown dwarf regime 
in the cloud complexes of the Gould Belt (cf. Andr\'e \& Saraceno 2005).
High-resolution ($0.01^{\prime\prime}-0.1^{\prime\prime}$) studies with 
the `Atacama Large Millimeter Array' 
(ALMA, becoming partly available in 2011, 
fully operational in 2013 -- cf. Bachiller 2008) 
at $\sim $~450~$\mu $m~--~3~mm will 
allow us to probe the kinematics of individual condensations in distant, massive protoclusters.
Complementing each other nicely, $Herschel$ and ALMA will tremendously 
improve our global understanding of the initial stages of star formation 
in the Galaxy.

%\begin{thereferences}{99}

%\begin{thebibliography}{99}

%
%\end{thereferences}

\end{document}